\documentclass[aps, pra, twocolumn, notitlepage,superscriptaddress   
]{revtex4-1}

\usepackage{amssymb,latexsym,amsmath}
\usepackage{bm}
\usepackage{amssymb,amsmath}
\usepackage{graphicx,xcolor}
\usepackage{epstopdf}

\usepackage{hyperref}   
\hypersetup{%
	pdfpagemode=None, 
	pdfstartpage=1,
	pdfmenubar=true,
	pdftoolbar=true,
	colorlinks = true,
	linkcolor=blue,
	citecolor=blue,
	urlcolor=blue,
	bookmarksopen=false
}

\newcommand{\beq}{\begin{equation}}   
\newcommand{\eeq}{\end{equation}}

\newcommand{\br}{{\bm r}}

\begin{document}

\title{Quantized superfluid vortex dynamics on cylindrical surfaces and planar annuli}
\author {Nils-Eric Guenther}
\email{nils.guenther@icfo.eu}
\affiliation{ICFO -- Institut de Ci\`encies Fot\`oniques, The Barcelona Institute of Science and Technology, 08860 Castelldefels (Barcelona), Spain}
\author {Pietro Massignan}
\email{pietro.massignan@upc.edu}
\affiliation{Departament de F\'isica, Universitat Polit\`ecnica de Catalunya, Campus Nord B4-B5, E-08034 Barcelona, Spain}
\affiliation{ICFO -- Institut de Ci\`encies Fot\`oniques, The Barcelona Institute of Science and Technology, 08860 Castelldefels (Barcelona), Spain}
\author {Alexander L.\ Fetter}
\email{fetter@stanford.edu}
\affiliation {Departments of Physics and Applied Physics, Stanford University, Stanford, CA 94305-4045, USA}
\date{\today}

\begin{abstract}
Superfluid vortex dynamics on an infinite cylinder differs significantly from that on a plane.  
The requirement that a condensate wave function be single valued upon once encircling the cylinder means that  such a single vortex cannot remain stationary.  
Instead, it acquires one of a series of quantized translational velocities around the circumference, the simplest being $\pm \hbar/(2MR)$, with $M$ the mass of the superfluid particles and $R$ the radius of the cylinder.
A generalization to a finite cylinder automatically includes these quantum-mechanical effects through the pairing of the single vortex and its image in either the top or bottom end of the surface. 
The dynamics of a single vortex on this surface provides a hydrodynamic analog of Laughlin pumping.
The interaction energy for two vortices on an infinite cylinder is proportional to the classical stream function $\chi({\bm r}_{12})$, and it crosses over from logarithmic to linear when the intervortex separation ${\bm r}_{12}$ becomes larger than the cylinder radius.
An Appendix summarizes the connection to an earlier study of Ho and Huang for one or more vortices on an infinite cylinder.
A second Appendix reviews the topologically equivalent  planar annulus, where such quantized vortex motion has no offset, but Laughlin pumping may be more accessible to experimental observation.
\end{abstract}

\maketitle

 \section{Introduction}\label{sI}

The dynamics of point vortices in an incompressible nonviscous  fluid  has been of great interest since the late 19th century~\cite{Lamb45}.  For example, given an initial vortex configuration, the subsequent motion obeys first-order equations of motion, which differs greatly from the usual second-order Newtonian  equations describing point masses.  In addition, the $x$ and $y$ coordinates of each vortex serve as canonical variables, analogous to $x$ and $p$ for a  Newtonian point particle.

This description has found wide application to superfluid $^4$He which acts  like an incompressible fluid for vortex motion much slower than the speed of sound $\sim 240$ m/s~\cite{Donn91}.  Such superfluid systems involve a complex macroscopic condensate wave function $\Psi = |\Psi| e^{i\Phi}$, whose phase $\Phi$ determines the superfluid velocity $\bm v = \hbar\bm \nabla \Phi/M$, where $M$ is the atomic mass.  In this way, the quantum-mechanical phase $\Phi$  becomes the velocity potential. The creation of dilute ultracold  superfluid atomic Bose-Einstein condensates (BECs) in 1995  has subsequently  stimulated many new applications of the same formalism~\cite{Pita03,Peth08, Fett09}.  

Classical nonviscous irrotational and incompressible hydrodynamics describes well the dynamics of  vortices in superfluid $^4$He, with the additional condition of quantized circulation~\cite{Donn91}.
Although dilute ultracold superfluid BECs are compressible, local changes in the density become small in the Thomas-Fermi (TF) limit, which typically describes many important experiments~\cite{Baym96}. 
In this limit, the condition of current conservation for steady flow  $\bm \nabla \cdot (n\bm v) = 0$ reduces to the condition of incompressibility $   \bm \nabla \cdot \bm v = 0$. For such incompressible flow, the stream function $\chi $ provides an important alternative description of the superfluid flow.  Specifically, for two-dimensional flow in the $xy$ plane, the velocity becomes 
\begin{equation}\label{vFlow}
\bm v = (\hbar/M)\,\hat {\bm n}\times \bm \nabla \chi,
\end{equation}
 where $\hat{\bm n} = \hat{\bm x}\times\hat{\bm y}$ is the unit vector normal to the two-dimensional plane, following the right-hand rule.

For such irrotational incompressible flow in two dimensions ($x,y$), the complex variable $z = x+iy$ provides a natural framework for vortex dynamics.  It is helpful to introduce a complex potential  $F(z) = \chi(\br) + i\Phi(\br)$, with $\br=(x,y)$. 
For any such analytic function,   the Cauchy-Riemann conditions yield the components of velocity:
\begin{equation}\label{CR}
v_x = \frac{\hbar}{M} \frac{\partial\Phi}{\partial x} = - \frac{\hbar}{M} \frac{\partial\chi}{\partial y}\quad\hbox{and}\quad v_y =  \frac{\hbar}{M} \frac{\partial\Phi}{\partial y} =  \frac{\hbar}{M} \frac{\partial\chi}{\partial x}.
\end{equation}
These conditions give the  compact representation of the hydrodynamic flow velocity components
\begin{equation}\label{vel}
v_y + iv_x = (\hbar/M)\,F'(z)
\end{equation} 
in terms of the first derivative $F'(z)$ of the complex potential.

Note that the representation of $\bm v$  in terms of the velocity potential $\Phi$ ensures that the flow is irrotational (namely $\bm \nabla \times \bm v=\bm 0$), apart from singularities associated with the vortex cores.  This applies to all superfluids in both two and three dimensions.  In contrast, the representation in terms of the stream function ensures that the flow is incompressible with $\bm \nabla\cdot \bm v = 0$, for it can be rewritten as $\bm v = -(\hbar/M)\bm\nabla\times (\hat {\bm n} \chi)$.  This condition does  not apply generally to all superfluids, but it can be very useful in many  specific cases.

For flow in  a plane, the stream function $\chi(\br)$   has the special advantage that $\chi$  takes a constant value along a streamline of the hydrodynamic flow.  
In addition, as shown below,  the interaction energy between two vortices at $\bm r_1$ and $\bm r_2$ is directly proportional to $\chi(\br_{12})$, where $\bm r_{12} =\bm r_1-\bm r_2$ is the intervortex separation, as discussed in~\cite{Cald17}.

All these results are familiar in the case of point vortices in a plane.  For example, the complex potential for a positive unit vortex at the origin is 
$F(Z) =\ln Z 
= \ln r + i\phi$, where $r=\sqrt{X^2+Y^2}$ and $\tan \phi = Y/X$. Hence $F'(Z) = 1/Z$.   It is not difficult to verify that $\bm v = (\hbar/Mr)\hat{\bm \phi}$ and that the vorticity  $\bm \zeta = \bm \nabla\times \bm v$ is singular, with $\bm \zeta = (2\pi \hbar/M)\, \hat{\bm n}\,\delta^{(2)}(\bm r)$.  It follows from familiar vector identities that the stream function for a point vortex at the origin obeys an inhomogeneous equation with the vorticity as its source:
\begin{equation}\label{2d}
\nabla^2 \chi(\br) = 2\pi \delta^{(2)}(\bm r)
\end{equation}
and is thus  effectively a two-dimensional Coulomb Green's function. In general, the stream function  also satisfies various boundary conditions.  Typically, boundaries break translational invariance, and the stream function depends symmetrically on the two variables: $\chi(\bm r,\bm r_j) = \chi(\bm r_j,\bm r)$.

More recently, the behavior of  singularities in various order parameters on curved surfaces has attracted great interest.  The simplest such case is a superfluid vortex with a single complex order parameter, although liquid crystals   present many more intricate examples~\cite{Lube92,Turn10}. 

Here we focus on the dynamics of point vortices in a thin superfluid film on a cylindrical surface of radius $R$.  
We start with an infinite cylinder in Sec.~\ref{secII} and show  that the identification of  the velocity potential  as the quantum-mechanical phase $\Phi$ requires a single vortex to move uniformly around the cylinder with (in the simplest case) one of two specific quantized values.  

In Sec.~\ref{sec.iii}  we study the dynamics of two vortices on a cylinder, which is unexpectedly rich.  Vortices  with opposite signs move uniformly perpendicular to the relative vector $\bm r_{12}$,  in the direction of the flow between them. 
This behavior is closely related to the quantized vortex velocity found in Sec.~\ref{secII}. In contrast, two such vortices with the same sign maintain their centroid $\bm R_{12} = \frac{1}{2}(\bm r_1 + \bm r_2)$, displaying both bound orbits and unbounded orbits, in  close analogy to the motion of a simple pendulum. 

In Sec.~\ref{inter},  we evaluate the interaction energy $E_{12}$ of two vortices, relating it to the relevant stream function $\chi(\bm r_{12})$.  This result allows us to re-express the dynamics of two or more vortices in terms of forces, including the Magnus force~\cite{Cald17}.  

Section~\ref{sec:finiteCylinder} considers a vortex on  a finite cylinder of length $L$, where the method of images provides an exact solution in terms of the first Jacobi $\vartheta$ function~\cite{Whit62,Fett67}.
The resulting dynamics under the action of additional external rotation constitutes a direct hydrodynamic analog of the Laughlin pumping.

Previously, Ho and Huang~\cite{Ho2015} studied spinor condensates on a cylindrical surface and found some of the results that we present here.  Appendix \ref{sec:conformal_transformation} compares the two approaches.

Appendix \ref{sec:annulus} reviews the annular geometry in a plane, considered in Ref.~\cite{Fett67}. The geometry of a planar  annulus is topologically equivalent to that of a finite cylinder, so that the vortex dynamics on these two surfaces have some close resemblances.

\section{Point vortex on an infinite cylinder}\label{secII}

On the surface of  a cylinder of radius $R$,  let $-\pi R\le x \le \pi R$ represent the coordinate around the circumference and $-\infty \le y \le \infty$ the unbounded coordinate along the cylinder's axis.  The unit vector $\hat{\bm n}$ is then the outward normal to the surface of the cylinder.  For a thin superfluid film, the problem is apparently equivalent to  the infinite plane with periodic repetitions of a strip of width $2\pi R$ along $\hat{\bm x}$.  As seen below, however, this classical picture violates the quantum-mechanical requirement of a single-valued condensate wave function on once encircling the cylinder.  We find that a single vortex on an infinite  cylinder must move around the cylinder with a set of quantized velocities.

Throughout this section we consider a single point vortex at the origin of the cylindrical surface ($x=0,y=0$).
Note that $x/R$ is  just the azimuthal angle $\phi$ in cylindrical polar coordinates.
With the usual complex notation $z = x+ i y$ and  $f(z)$ a function of this complex variable, the complex potential  $F(z) = \ln f(z)$ corresponds to a positive vortex at each zero of $f(z)$.  In particular, Sec.~156 of~\cite{Lamb45} notes that the choice 
\beq
F(z) = \ln\left[\sin\left(\frac{z}{2R}\right)\right]
\eeq
represents a  one-dimensional periodic array of positive vortices at positions $z_n = 2\pi n R$, with $n\in \mathbb{Z}$.   

 At first sight, this complex potential should also represent a single superfluid vortex at the origin of an infinite cylinder.  Note, however, that $\sin(z/2R)$ changes sign for $z \to z+2\pi R$.  Consequently,  the present velocity potential  $\Phi(\bm r)  = \Im F(z) $ is not acceptable as the phase of a single-valued quantum-mechanical condensate wave function,  because $e^{i\Phi(\bm r) }$  remains unchanged  only for   $x \to x+ 4\pi R$.

\subsection{Velocity potential for one vortex on a cylinder:  classical hydrodynamics {\it vs.} quantized superfluidity}\label{quantized}

Here, we explore the source of this discrepancy by evaluating in detail the velocity potential
\beq
\Phi (\br) = \Im \ln\left[\sin\left(\frac{z}{2R}\right)\right]
=\arctan\left[\frac{\tanh\left(y/2R\right)}{\tan\left(x/2R\right)}\right]. 
\eeq
As expected, this reduces to $\arctan(y/x)$ for $r\ll R$.

In the quantum interpretation, the velocity potential $\Phi$ is also the phase of the condensate wave function, which leads to the following inconsistency:  Note that  $\Phi$ increases by $\mp{\rm sgn}(y)\pi$  when $x\to x+2\pi R$, where ${\rm  sgn}(y) = |y|/y$. Hence the condensate wave function would be antiperiodic on once going around the cylinder.  As seen below, the actual fluid velocity itself is indeed continuous and periodic, so that this complex potential is  acceptable as a classical solution but not as a quantum mechanical one if the vortex itself remains at rest.

\begin{figure*}[t]
\begin{center}
     \includegraphics[width=\linewidth]{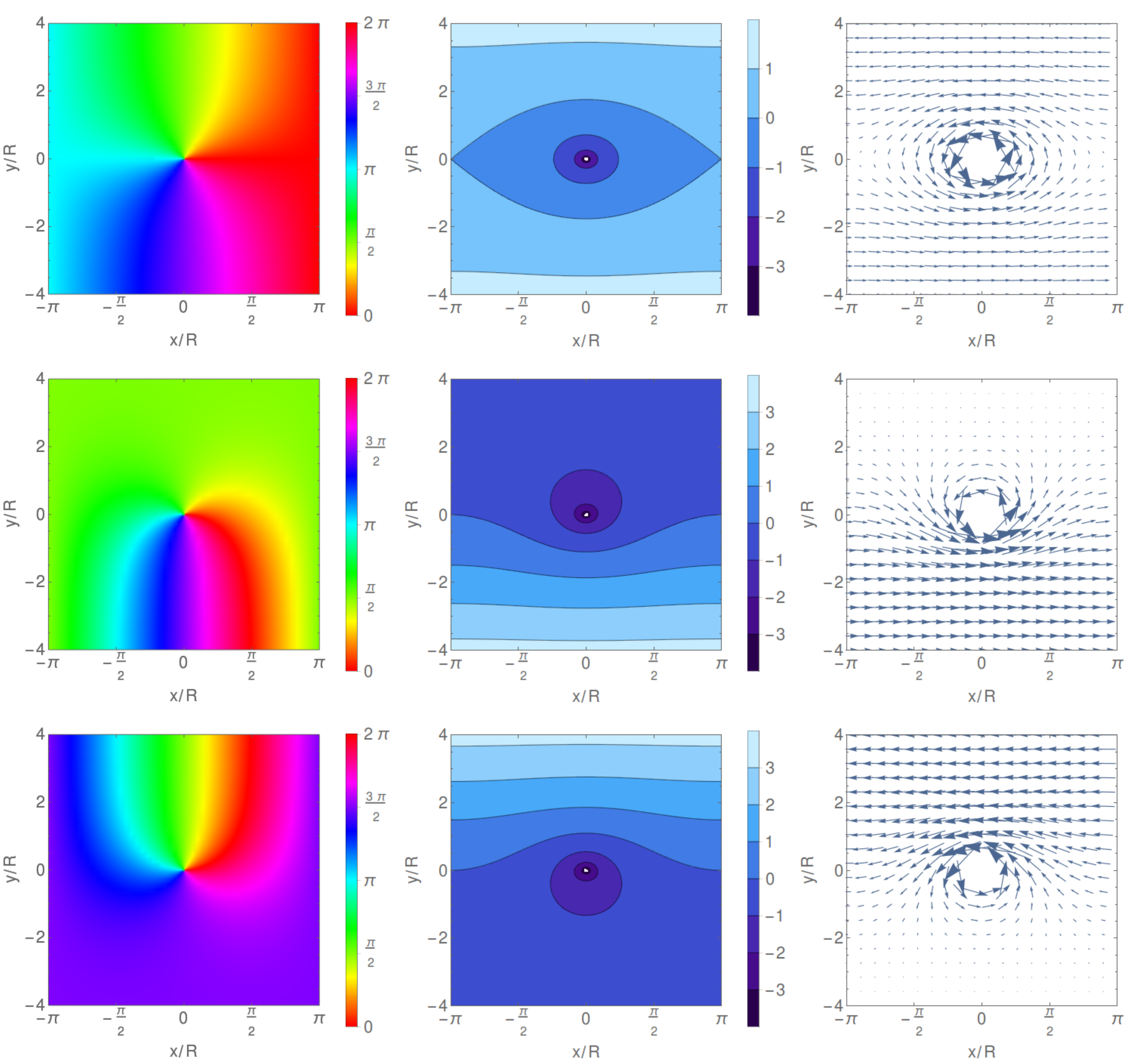}
\caption{\label{fig:singleVortexInfiniteCyl}
     {Phase (left), streamlines (center), and velocity flow (right) for single point vortex on an infinite cylinder}. Top:~No additional uniform flow; 
     since the phases at $x=\pm \pi/R$ differ by $\pi$, this is not an acceptable solution for a quantum superfluid.
     Middle and bottom:~Additional uniform flow specified by $C=1/2$ and $C=-1/2$, respectively (or, equivalently, $n_\uparrow=0$ and $n_\uparrow=-1$)}.
\end{center}
\end{figure*}

Consider the lines of constant phase.  For a single vortex on a plane, these lines extend radially from the center of the vortex, rather like electric field lines from a two-dimensional point charge.  On a cylinder, the periodicity means that half the phase lines go upward and half go downward (see Fig.~\ref{fig:singleVortexInfiniteCyl} top row).  The net change in phase on going once around the circumference of the cylinder is $\mp \pi$, depending on the sign of $y$.

An illuminating way to think about this question  focuses on a vorticity flux of $2\pi$ associated with a singly quantized vortex (measured in units of $\hbar/M$).  For a positive stationary vortex with charge $q=1$ at the origin of the surface, the flux comes from inside the cylinder and emerges radially  outward along $\hat{\bm n}$ through the center of the vortex.
In the present case of a vortex at rest, the flux comes symmetrically with $\pi$ flowing downward from $y\to \infty$ and $\pi$ flowing upward from $y\to -\infty$, as is clear from Fig.~\ref{fig:singleVortexInfiniteCyl} top row.  
Physically, the vortex on the surface can be considered the end of a vortex line in a superfluid filling the interior of the cylinder.  
Clearly the vortex line must come wholly from one end or the other, so that such a flux splitting is not possible.  We'll see that a moving vortex indeed satisfies these conditions (for special values of the motion).
 
More generally, the flow velocity is $\bm v = \hbar\bm \nabla \Phi/M$.  Hence adding a uniform flow  in the $\hat{\bm x}$ direction will alter the phase gradient and thus alter the phase change in going around the cylinder.  Specifically, consider the more general complex potential for a vortex located at $x_0=0$ on the cylinder
\beq\label{C}
F(z) = \ln \left[\sin\left(\frac{z}{2R}\right)\right] +i \,C\frac{z}{R},
\eeq
where $C$ is a dimensionless real constant. The additional uniform velocity is $(\hbar/MR) \, C\hat{\bm x} $.  As a result, the previous expression for $\Phi$ acquires the additional term $ C x/R$.  The  additional net phase change on going once around the cylinder is $2\pi C$.  Thus merely adjusting the value $C$ can yield any desired phase change;  for example, the choice $C =1/2$ would give zero total phase change for $y>0$ and $2\pi$ total phase change for $y<0$.  In essence, this behavior is simply the hydrodynamic analog of the familiar Bohm-Aharonov effect.  Alternatively, it represents a sort of gauge freedom to alter the phase.  
 
Figure \ref{fig:singleVortexInfiniteCyl}(middle and bottom) shows the phase pattern for a single positive point vortex with  additional flow velocity $C = \pm 1/2$.  These choices ensure that the lines of constant phase all collect into the lower (upper) part of the cylinder for $C = 1/2\,(-1/2)$ leaving the fluid asymptotically at rest in the upper (lower) end of the cylinder.  In these special cases, the net change in phase upon once encircling the cylinder now will be $0$ or $\pm2\pi$, merely by counting the phase lines crossing the path.  Note that the two solutions may be mapped onto each other by a rotation of 180$^\circ$, which effectively interchanges the two ends of the cylinder. 
 Evidently, quantum mechanics requires that the phase lines from a single vortex on a cylinder  must flow to $\pm \infty$ in multiples of $2\pi$ to ensure that the condensate wave function is single valued.
  
To clarify these questions, consider a single positive  vortex on a cylinder at the origin.  Integrate in a positive sense the hydrodynamic fluid velocity around a closed rectangular contour (namely the circulation) with vertical sides at $x=\mp \pi R$.  Clearly, the contributions of these vertical sides cancel because of the periodicity of the flow velocity.  In addition, the circulation integral will be $0$ if the horizontal parts do not enclose the vortex,  and $2\pi$ if the horizontal parts  do enclose the vortex.  For $C=0$ (namely no external flow), the top and bottom parts each contribute $\pi$ to the dimensionless circulation.  If $C\neq 0$, the contributions of the top and bottom parts each shift linearly in such a way that the net circulation is unchanged.  In particular,  for 
$C = n_\uparrow+\frac{1}{2}$
 with $n_\uparrow$ an integer that specifies the quantum of circulation on the horizontal path above the vortex, each horizontal part contributes a multiple of $2\pi$.  

The additional uniform flow  means that the vortex now moves uniformly around the cylinder with quantized velocity, required to satisfy the quantum-mechanical condition that the condensate wave function be single valued.
 For any complex velocity function $F'(z)$ that contains a vortex at some point $z_0$, the following  limit gives  the complex velocity of that vortex as
\begin{equation}\label{vortex_velocity}
\bm{\dot}y_0 +i\bm {\dot}x_0= \frac{\hbar}{M}  \lim_{z\to z_0}\left[ F'(z) -\frac{1}{z-z_0}\right]. 
\end{equation}
In the present case, this expression simply reproduces the previous result that the vortex moves with the local uniform flow velocity.  For $C= 0$  with no applied flow, any particular vortex remains at rest, either from this mathematical treatment or more physically by noting that the induced flow at any particular vortex cancels because of the left-right symmetry of the one-dimensional periodic array.

Focus on the two simplest cases with $C=\pm 1/2$, in which case the flow vanishes as $y\to  \pm\infty$ [see Fig.~\ref{fig:singleVortexInfiniteCyl}~(middle and bottom)].  The corresponding complex potential becomes 
\beq\label{HH}
F_\pm(z) = \ln\left[\sin \left(\frac{z}{2R}\right)\right] \pm \frac{i z}{2R} = \ln\left(e^{\pm iz/R}-1\right) 
  + {\rm  const}.
\eeq
Apart from the additive constant, this complex potential is just that considered by Ho and Huang~\cite{Ho2015} as the two possible conformal transformations from a plane to a cylinder (corresponding to the choice $\pm i$). This connection clarifies the special role of the two values $C = \pm 1/2$.  We consider this point in detail in Appendix~\ref{sec:conformal_transformation}. 

\subsection{Stream function for one vortex on a cylinder}\label{ssII.2}

As noted in Sec.~\ref{sI}, the stream function $\chi(\bm r) = \Re F(z)$ provides a clear picture of the hydrodynamic flow through its contour plots.  In the present case, $\chi(\bm r)$ 
 is a little intricate, which illustrates a principal advantage of this complex formalism.   Specifically, 
the stream function for one vortex on the surface of a cylinder of radius $R$ is
\begin{align}\label{chiAlt}
\chi(\bm r) &= \Re \left\{\ln\left[\sin\left(\frac{z}{2R}\right)\right] +iC\frac{z}{R}\right\}\nonumber\\
&= \frac{1}{2} \ln\left |\sin\left(\frac{x+iy}{2R}\right)\right|^2 - C\,\frac{y}{R}.  
\end{align}
Familiar complex trigonometric identities give 
\begin{align}\label{chi1}
\chi(\br) &= \frac{1}{2} \ln\left[\sin^2\left(\frac{x}{2R}\right) + \sinh^2\left(\frac{y}{2R}\right) \right] - C \,\frac{y}{R}\nonumber\\
&=\frac{1}{2} \ln\left[\frac{1}{2}\cosh\left(\frac{y}{R}\right) -\frac{1}{2} \cos\left(\frac{x}{R}\right) \right] - C \,\frac{y}{R},
\end{align}
where each form is useful in different contexts.

This stream function  has the proper periodicity in $x$ and reduces to the result  
$ \frac{1}{2} \ln[(x^2 + y^2)/4R^2] -Cy/R= \ln (r/2R)-Cy/R$
 for a single vortex at the origin when  $r\ll R$. 
In contrast, for $|y| \gg R$, the stream function has the very different and asymmetric behavior $\chi(\br) \approx |y|/(2R) - Cy/R$, independent of $x$.    Correspondingly, $\bm\nabla \chi(\br) \approx  \hat{\bm y}\,[{\rm sgn}(y)/(2R) - C/R]$ in this limit, and the hydrodynamic flow velocity reduces to  a uniform flow (from $C$)  plus an antisymmetric uniform flow: $\bm v(\br) = (\hbar/M)\,\hat{\bm n}\times \bm \nabla \chi(\br)   \approx -\hat{\bm x}\,[(\hbar/2MR)\,{\rm sgn}(y)-\hbar C/MR]$.
  
To understand this asymptotic behavior, consider the induced flow of the corresponding infinite one-dimensional array of positive vortices in the plane (for simplicity, take  $C=0$). Close to each vortex, the flow circulates around that vortex in the positive sense, but for $|y|\gtrsim \pi R$, the combined flow instead resembles that of a ``vortex sheet'' (see Sec.~151 of~\cite{Lamb45}).    
Specifically, a vortex sheet arises when the transverse velocity has a discontinuity.  For example, consider the antisymmetric  flow field $\bm v =-v_0 \,\hat{\bm x}\, {\rm sgn} (y) $. Here, the vorticity is  $\bm \nabla  \times \bm v =2 v_0 \,\delta (y)\, \hat{\bm z} $, which follows either by direct differentiation or with Stokes's theorem.
In particular,  the asymptotic  flow from a periodic array of positive unit  vortices along the $x$ axis with spacing $2\pi R$ approximates a vortex sheet with $v_0 =\hbar/(2MR)$.

Evidently, the hydrodynamic flow for a single vortex on a cylinder is considerably more complicated than that for a single vortex in the plane.  Note that the hydrodynamic  flow arises from an analytic function $F(z)$, so that $\chi(\bm r)$ and $\Phi(\bm r)$ both  satisfy Laplace's equation (apart from the local singularity associated with the vortex). Such a two-dimensional function cannot be periodic in both directions.  Instead, the sum of the  curvatures associated with $x$ and $y$ must vanish, so that the solution necessarily decays exponentially for large $|y|$ (in this case, to a nonzero constant), as seen here from the  hyperbolic functions in $\Phi(\bm r)$ and $\chi(\bm r)$.
 
In Fig.\ \ref{fig:singleVortexInfiniteCyl}(top middle) we show a contour plot of the stream function $\chi(\br)$ for a single vortex on the surface of a cylinder with $C=0$. 
Lamb~\cite{Lamb45} has a similar figure in Sec.~156.
As expected, the  streamlines on  the cylindrical surface exhibit both the  periodicity in $x$ and the exponential decay of the motion in the $\pm y$ direction with the characteristic length $\sim R$.  
Note the occurrence of two topologically different types of trajectories. This phase plot  resembles  that of a simple pendulum, which reflects the similar canonical  roles of $x,y$ for a vortex and $x,p$ for a pendulum. Here, the separatrix is parametrized by $\chi(\br) = 0$, namely by $\sin^2(x/2R) + \sinh^2(y/2R) = 1$. Inside the closed curves of the separatrix, the flow circulates around the vortex and its periodic images.  Outside the separatrix, the flow continues in one direction, like a pendulum with large energy.    In the present hydrodynamic context,  streamlines inside the separatrix  correspond to ``libration'' of the pendulum and encircle the vortex with  zero winding number around the cylinder.  Otherwise,  streamlines correspond to ``rotation'' of the pendulum.  They do  not encircle the vortex but have winding number $\mp 1$ around the cylinder, depending on the value of $-{\rm sgn}(y)$.

With standard trigonometric identities, it is not hard to find the hydrodynamic flow velocity induced by the single positive vortex at the origin on the surface of a cylinder (here, for simplicity, we take $C=0$): 
\begin{align}\label{v_cyl}
\bm v(\br) &= \frac{\hbar}{2MR}  \,\frac{-\hat{\bm x}\sinh (y/R)+\hat{\bm y}\sin (x/R)}{\cosh( y/R)-\cos (x/R)} \nonumber\\
&= \frac{\hbar}{2MR}\,\hat{\bm n}\times \left[\frac{\hat{\bm x} \sin (x/R) + \hat{\bm y} \sinh (y/R) }{\cosh (y/R)-\cos (x/R)}\right] \nonumber\\
&= \frac{\hbar}{M}  \,\hat{\bm n} \times \bm \nabla \chi(\br).
\end{align}
The resulting flow pattern is shown in Fig.~\ref{fig:singleVortexInfiniteCyl}(top right).
For $|x|\ll R$ and $|y|\ll R$, the hydrodynamic flow field  reduces to the familiar expression $\bm v(x,y) \approx (\hbar/M)\,\hat{\bm n}\times \bm r/r^2 = (\hbar/Mr)\,\hat{\bm \phi}$, which falls off inversely with the distance from the vortex in all directions.
For large $|y|/R$  on a cylinder, in contrast, the flow velocity  reduces to a constant $\bm v(x,y)\approx -(\hbar/2MR)\,\hat{\bm x}\, {\rm sgn}(y)$.
In this region the periodicity around the cylinder dominates the flow pattern, rather than the single vortex.

\section{Multiple vortices on a cylinder}\label{sec.iii}

It is now straightforward to generalize the previous discussion to the case of $N$ vortices on an infinite cylinder, each located at complex position $z_j$ and with charge $q_j = \pm 1$ ($j=1,\ldots,N$). 
As in electrostatics, the complex potential  for multiple vortices on the cylindrical surface is simply the sum of the complex potentials of the individual vortices, always with the option of adding a uniform flow of the form  $iCz/R$.  For an even number of vortices, however, this term is unnecessary. 
\begin{equation}
 F^{(N)}(z) = \sum_{j=1}^N q_jF(z-z_j)
= \sum_{j=1}^N q_j\ln\left[\sin\left(\frac{z-z_j}{2R}\right) \right].
 \end{equation}
The corresponding velocity potential $\Phi^{(N)}(\bm r)$ and stream function $\chi^{(N)}(\bm r)$ are the imaginary and real parts of $F^{(N)}(z) $ and need not be given explicitly.

\subsection{Induced motion of two vortices on a cylinder}
\label{sec:induced_motion_of_two_vortices}

As noted at the end of Sec.~\ref{quantized}, in the absence of external flow a single vortex on a cylinder remains stationary.  Consequently,  the motion of each vortex arises only from the presence of the other vortex.  Equation (\ref{vortex_velocity})  immediately gives  the complex velocity of the first vortex 
\begin{equation}
\bm{\dot}y_1 + i \bm{\dot}x_1= \frac{\hbar}{MR} \frac{ q_2}{2}\,\cot\left(\frac{z_1-z_2}{2R}\right),
\end{equation}
 and similarly 
\begin{equation}
\bm{\dot}y_2 + i \bm{\dot}x_2 
= -\frac{\hbar}{MR} \frac{ q_1}{2}\,\cot\left(\frac{z_1-z_2}{2R}\right).
\end{equation}

It is now helpful to introduce the vector notation used in Sec.~\ref{quantized}.
For two vortices at $\bm r_1$ and $\bm r_2$, let $\bm R_{12} = \frac{1}{2} (\bm r_1 + \bm r_2)$ be the centroid and $\bm r_{12} = \bm r_1-\bm r_2$ be the relative position (note that the vector  $\bm r_{12}$ runs from 2 to 1).
As a result [compare Eq.~\eqref{v_cyl}], we find the appropriate dynamical equations 
\begin{equation*}
 \bm{\dot R}_{12} = \frac{\hbar}{MR}\left(\frac{q_1-q_2}{4}\right)\,\left[ \frac{\hat{\bm x} \sinh (y_{12}/R) -\hat{\bm y}\,\sin (x_{12}/R)}{\cosh (y_{12}/R) -\cos (x_{12}/R)}\right] , 
 \end{equation*}
and 
\begin{equation*}
 \bm{\dot r}_{12} = \frac{\hbar}{MR}\left(\frac{q_1+q_2}{4}\right)\,\left[ \frac{-\hat{\bm x} \sinh (y_{12}/R) +\hat{\bm y}\,\sin (x_{12}/R)}{\cosh (y_{12}/R) -\cos (x_{12}/R)}\right].
 \end{equation*}

\subsection{Two vortices with opposite signs (vortex  dipole)}

When $q_1=1$ and $q_2=-1$, the vortex dipole moves with no internal rotation $\bm{\dot r}_{12} =\bm 0$ (so that $x_{12}$ and $y_{12}$ remain constant,  simplifying the subsequent dynamics).  Furthermore, the centroid moves with uniform translational velocity 
\begin{align}
  \bm{\dot R}_{12} &= \frac{\hbar}{2MR}\,\left[ \frac{\hat{\bm x} \sinh (y_{12}/R) -\hat{\bm y}\,\sin (x_{12}/R)}{\cosh (y_{12}/R) -\cos (x_{12}/R)}\right] \nonumber\\
  &=- \frac{\hbar}{M} \,\hat{\bm n} \times \bm \nabla \chi(\bm r_{12}).
  \end{align}
at fixed $x_{12}$ and $y_{12}$. Several limits are of interest:  
\begin{enumerate}
\item If $|x_{12}|\ll R$ and $|y_{12}|\ll R$, then the translational velocity is the same as that for a vortex dipole on a plane: 
\begin{equation}
\bm{\dot R} _{12}= \frac{\hbar}{M} \frac{\hat{\bm x}\, y_{12}- \hat{\bm y}\,  x_{12}}{x_{12}^2 + y_{12}^2}    
= -\frac{\hbar}{M}  \frac{\hat{\bm n}\times\bm r_{12}}{r_{12}^2} . 
\end{equation}
Detailed analysis confirms that the vortex dipole moves in the direction of the flow between their centers.
\item If $|y_{12}|\gg R$, then the ratio of hyperbolic functions leaves only the $\hat{\bm x}$ component, with $\bm{\dot R}_{12} = (\hbar/2MR)\,\hat{\bm x}\;{\rm sgn}(y_{12})$.  This value reflects the hydrodynamic flow from a periodic array of vortices, as mentioned near the end of Sec.~\ref{ssII.2}.  In this limit, the vortex dipole will circle the cylinder in a time $4\pi M R^2/\hbar$.  Note that this motion is the same as that induced for one vortex with $C=\pm 1/2$, discussed in Sec.~\ref{quantized}.  Hence the additional induced motion of a single vortex on an infinite cylinder  can alternatively be thought to arise from a phantom negative vortex placed at $y_0' \to \mp \infty$, corresponding to $C = \pm 1/2$.

\end{enumerate}

\begin{figure*}[t]
\begin{center}
    \includegraphics[width=\textwidth]{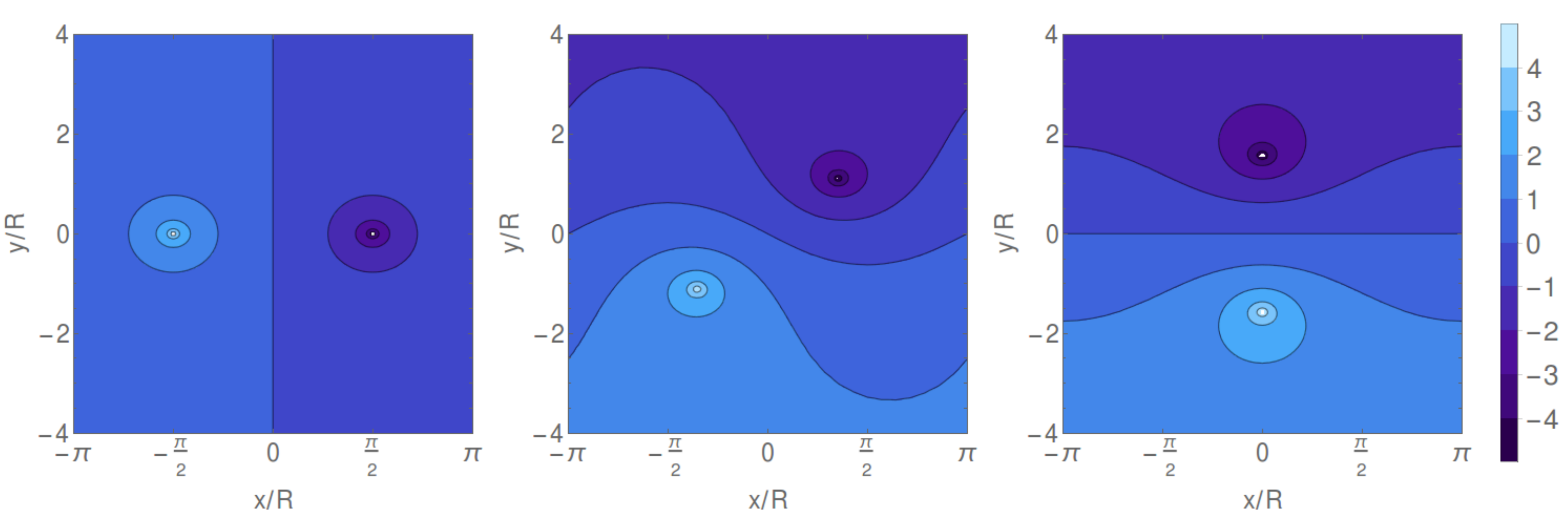}
     \caption{\label{fig:vortex_dipole}
Hydrodynamic streamlines for vortex dipole $q_1= - q_2 = 1$ on a cylinder of radius $R$, for various orientations of the dipole axis.}
 \end{center}
\end{figure*}

As an example, Figure~\ref{fig:vortex_dipole} shows the hydrodynamic streamlines for various orientations of the relative position $z_1-z_2$.
Specifically,  we plot the corresponding stream function $\chi$ for three typical cases:  $y_{12} =0$ with motion along $-\hat{\bm y}$, $x_{12}= y_{12}$, with motion along $\hat{\bm x} - \hat{\bm y}$, and $x_{12} = 0$ with motion along $\hat{\bm x}$. 

\subsection{Two vortices with same signs}

In the case of two positive vortices ($q_1=q_2 = 1$), it follows immediately that $\bm {\dot R}_{12} $ vanishes, so that the centroid of the two vortices remains fixed.  In contrast, the relative vector obeys the nontrivial equation of motion 
\begin{align}
  \bm{\dot r}_{12} &= \bm{\dot}x_{12} \,\hat{\bm x} + \bm{\dot}y_{12} \,\hat{\bm y}\nonumber\\
&=\frac{\hbar}{M}\,\left[ \frac{-\hat{\bm x} \sinh (y_{12}/R) +\hat{\bm y}\,\sin (x_{12}/R)}{\cosh (y_{12}/R) -\cos (x_{12}/R)}\right]\nonumber\\
&=\frac{\hbar}{M} \hat{\bm n}\times \bm\nabla \chi(\bm r_{12}) = \bm v(\bm r_{12}). 
\end{align}
Unlike the case of opposite charges (where $\bm{\dot r}_{12}$ remains fixed), the relative vector $\bm r_{12}$ now becomes time dependent. In fact, the last form given above shows that the motion of the two positive vortices precisely follows the hydrodynamic flow velocity of a single vortex $\bm v(x_{12},y_{12})$.  Hence the streamlines in Fig.~\ref{fig:singleVortexInfiniteCyl} (central column) completely  characterize  the  motion.  Several cases are of interest.
\begin{enumerate}
\item If $x_{12}^2 + y_{12}^2\ll R^2$, then the two positive vortices simply circle in the positive sense around their common center  $\bm R_{12}$.  The curvature of the surface is irrelevant and the motion is the same as on a flat plane.
\item  If $\sin^2(x_{12}/2R) + \sinh^2(y_{12}/2R) < 1$, the two vortices execute closed orbits in the positive sense around their common center $\bm R_{12}$, but the general orbits are not circular [by definition, they remain inside the separatrix in  Fig.~\ref{fig:singleVortexInfiniteCyl} (top row, central column)].
\item  If $\sin^2(x_{12}/2R) + \sinh^2(y_{12}/2R) > 1$, the two vortices move in opposite directions,  executing periodic closed orbits around the cylinder with unit winding number.  The upper vortex moves  monotonically to the left  and the lower vortex moves monotonically to the right, as seen in Fig.~\ref{fig:singleVortexInfiniteCyl} (top row, central column)  (they remain outside the separatrix).
\item For relatively large $|y_{12}|/R$, an expansion of the above equation yields the approximate form
\begin{multline}
\bm{\dot r}_{12} \approx \frac{\hbar}{MR} \left\{ -{\rm sgn}(y_{12}/R)\left[1-2\cos (x_{12}/R)\,e^{-|y_{12}|/R}\right]\,\hat{\bm x} \right.\\ 
\left.+ 2\sin (x_{12}/R)\, e^{-|y_{12}|/R}\, \hat{\bm y}\right\}.
\end{multline}
Asymptotically for $|y_{12}|\gg R$, the variable $x_{12}$ varies linearly in time.  The leading correction to this 
 uniform horizontal motion is a small periodic modulation  for  both $\hat{\bm x}$ and $\hat{\bm y}$ components.
\end{enumerate}

\section{Energy of two vortices}\label{inter}

The stream function $\chi(\bm r)$ provides the hydrodynamic flow velocity $\bm v(\bm r)$ through Eq.~(\ref{vFlow}), which is its usual role.  As shown below, however, the stream function also determines the interaction energy $E_{12}$  between two point vortices through Eq.~(\ref{E12}).  The analogous electrostatic situation is familiar in that the electrostatic potential gives both the electric field from a single point charge and the interaction energy of two point charges.  For electrostatics, this connection follows directly as the work done to bring the second charge in from infinity.  For vortices, however, such an argument is less clear, since vortices do not act like Newtonian particles and obey first-order equations of motion.  
Hence we present a straightforward analysis that gives the interaction energy $E_{12}$ of two vortices by integrating the kinetic-energy density, which is proportional to the squared velocity field.  This approach is clearly analogous to finding the electrostatic interaction energy of two charges by integrating the electrostatic-energy density, which is proportional to the squared electrostatic field.

In the present model, the total energy of two vortices at $\bm r_j$ ($j=1,2$) with unit charge $q_j=\pm1$ is the  spatial integral of the kinetic-energy density 
\begin{eqnarray}
E_{\rm tot}& =& \textstyle{\frac{1}{2}} nM \int d^2 r  \left[q_1\,\bm v(\bm r-\bm r_1) + q_2\,\bm v(\bm r-\bm r_2)\right]^2\nonumber\\[.2cm]
&=& \textstyle{\frac{1}{2}} nM \int d^2 r \left[ |\bm v(\bm r-\bm r_1)|^2+|\bm v(\bm r-\bm r_2)|^2\right.\nonumber\\[.2cm]
 &&\left.+ 2q_1q_2\,\bm v(\bm r-\bm r_1)\cdot\bm v(\bm r-\bm r_2)\right]
\end{eqnarray}
over the surface of the cylinder.  Here, $\bm v(\bm r)$ is the hydrodynamic velocity field of a single positive unit vortex at the origin, $n$ is the two-dimensional number density, and $M$ is the atomic mass.

\subsection{Interaction energy}
As noted in Secs.~\ref{quantized} and \ref{ssII.2}, for large $|y|/R$, the asymptotic velocity field of a single vortex on a cylinder is uniform.  Hence the kinetic energy of any single vortex diverges linearly as the upper and
lower integration boundaries on the cylinder become large (namely $|y|=Y\to \infty$).  As a result, each term in the above  kinetic energy of two vortices on a cylinder separately diverges.  The only case with a finite total kinetic energy is the vortex dipole with (say) $q_1 = 1$ and $q_2=-1$, since the two asymptotic hydrodynamic velocity flow fields then cancel.

It is convenient to use the stream function to characterize the local fluid velocity of the $j$th vortex:  $\bm v(\bm r-\bm r_j) =(\hbar/M)\hat{\bm n}\times \bm\nabla \chi_j$, where $\chi_j =\chi(\bm r-\bm r_j)$ [compare Eq.~(\ref{vFlow})].  The operation $\hat{\bm n}\times$ simply rotates the following vector through $\pi/2$ and  we find 
\begin{eqnarray}\label{intbyparts}
E_{\rm tot}& =& \textstyle{\frac{n\hbar^2}{2M}} \int d^2 r  \left(q_1 \bm \nabla \chi_1 +q_2\bm \nabla \chi_2\right)^2 \\[.2cm]\nonumber
&=& \textstyle{\frac{n\hbar^2}{2M}} \int d^2 r \left\{\bm \nabla\cdot\left[\left(\chi_1 + q_1q_2\chi_2\right)\bm \nabla\left(\chi_1 + q_1q_2\chi_2\right)\right] \right.\\[.2cm]\nonumber
&  & - \left.\chi_1\nabla^2\chi_1-\chi_2\nabla^2\chi_2 -q_1q_2 \left(\chi_1\nabla^2 \chi_2 +\chi_2\nabla^2\chi_1\right) \right\}.
\end{eqnarray}
We follow de Gennes's  argument for type-II superconductors~\cite{dege66}, but the analysis is also familiar from classical electrostatics. 
Here, the two-dimensional surface integral runs over the region $-\pi R\le x\le \pi R$ and $-Y\le y\le Y$, where $Y\to \infty$. 

The first term above involves the divergence of the  total derivative $\frac{1}{2}\bm \nabla \left(\chi_1+ q_1q_2\chi_2\right)^2$, and the divergence theorem reduces it to an integral on the boundary with outward unit normals.  The contributions from  the vertical parts at $x=\pm \pi R$ cancel because the integrand is periodic with period $2\pi R$.  In general, the contributions from the horizontal parts at $y = \pm Y$ separately diverge linearly, except for the special case of a vortex dipole with $q_1q_2 = -1$.  
The relevant quantity is $\frac{1}{2} \partial _y (\chi_1-\chi_2)^2$ for large $|y|$.  Equation~(\ref{chi1}) gives (here we take $C=0$ since the system is neutral)
\begin{eqnarray}
\chi_1-\chi_2 &=& \frac{1}{2}\ln\left[ \frac{\sin^2[(x-x_1)/2R] +\sinh^2[(y-y_1)/2R]}{\sin^2[(x-x_2)/2R] + \sinh^2[(y-y_2)/2R]}\right]\nonumber\\[.2cm]
&\approx& \frac{|y-y_1|-|y-y_2|}{2R}+\cdots\nonumber\\[.2cm]
&=& {\rm constant} + \cdots \ {\rm for}\ |y| \to \infty,
\end{eqnarray}
where the corrections are exponentially small for large $|y|$.
It is now clear that each  horizontal contribution  vanishes for the present case of a vortex dipole,  reflecting the overall charge neutrality.  

It remains to evaluate the second line of Eq.~(\ref{intbyparts}).  We already noted that $\nabla^2\chi_j= \nabla^2 \chi (\bm r - \bm r_j) =  2\pi \delta^{(2)}(\bm r-\bm r_j)$, and  the interaction energy (the terms involving the cross product of $\chi_1$ and $\chi_2$) thus becomes 
\begin{equation}\label{E12}
E_{12} = - (2\pi n\hbar^2/M)\,q_1q_2 \,\chi(\bm r_{12}),
\end{equation}
 for general choice of $q_1q_2$.
The dynamics of two vortices  involves  gradient operations like  $\bm \nabla_1 E_{12}$, so that any divergent constant becomes irrelevant  (alternatively, we can redefine the zero of the energy).
As in Eq.~(7) of~\cite{Cald17},  it is convenient to take out a factor $2\pi n\hbar$, writing 
\begin{equation}\label{Vij}
V_{12} = -q_1q_2\,(\hbar/M) \,\chi(\bm r_{12}), 
\end{equation}
which properly reduces to $ q_1q_2\,(\hbar/M) \ln(2R/r_{12})$ for small intervortex separation.
 
\subsection{Self energy of one vortex}
\begin{figure}[t]
\begin{center}
\includegraphics[width=0.9\columnwidth,angle=0,clip=]{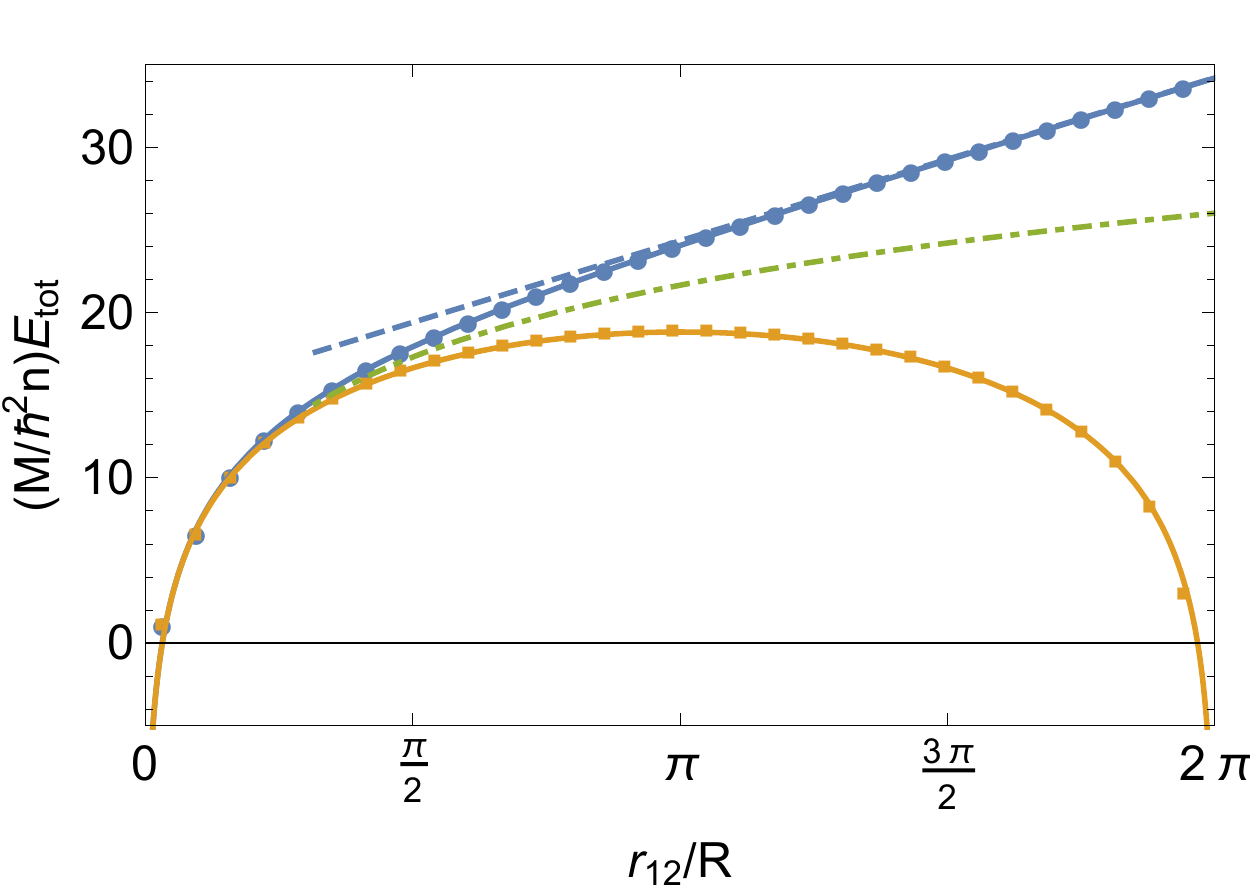}
\end{center}
\caption{\label{fig:VortexDipoleEnergy}
Energy of a vortex dipole on an infinite cylinder, as a function of the intervortex separation $r_{12}$, for a dipole oriented along the axis of the cylinder and moving along the equatorial direction (dark blue), and for a vortex dipole oriented along the equator and moving along the axial direction (light orange). 
The vortex core size is set to $\xi=R/10$. 
Symbols display the numerical evaluation of Eq.~\eqref{intbyparts}, and solid lines show the analytical result in Eq.~\eqref{EtotDipole}. 
The dot-dashed green line is the usual result on the plane, $E_{\rm tot}=(2\pi\hbar^2n/M)[\ln(r_{12}/2R)+\ln(2R/\xi)]$, and the dashed line is the asymptotic behavior for large axial separation $y_{12} \gg R$, $E_{\rm tot}=(2\pi\hbar^2n/M)[y_{12}/(2R)+\ln(R/\xi)]$.}
\end{figure}

Equation (\ref{intbyparts}) also contains two self energy terms, one for each vortex.  Consider a single vortex $1$ at the origin with self energy 
\begin{align}\label{E1}
E_1 &= \frac{n \hbar^2}{2M} \int d^2 r\,\bm\nabla \chi(\bm r)\cdot\bm\nabla \chi(\bm r) \nonumber\\
&= \frac{n\hbar^2}{2M}  \int d^2 r\,\left\{ \bm\nabla\cdot\left[ \chi(\bm r)\bm\nabla \chi(\bm r)\right] - \chi(\bm r)\nabla^2\chi(\bm r)\right\}.
\end{align}
A heuristic approach for the self energy terms (those involving $-\chi_j\nabla^2\chi_j$) in Eq.~(\ref{intbyparts})  is to cut off the singularity at the small  core radius $\xi$, which gives
 \begin{equation}\label{E11}
 E_1 =  \frac{\pi n \hbar^2}{M}\ln\left(\frac{2R}{\xi}\right).
 \end{equation}
 The finite total energy of a vortex dipole is simply the sum of the interaction energy and the two self energies 
 \beq\label{EtotDipole}
 E_{\rm tot}  = E_{12} + 2 E_1=  \frac{2\pi n\hbar^2}{M} \left[\chi(\bm r_{12}) + \ln\left(\frac{2R}{\xi}\right) \right].
 \eeq
 Note that this total vortex energy reduces to the familiar $\ln(r_{12}/\xi)$ for small $r_{12}$.  Otherwise it has a very different form and grows linearly for $|y_{12}|\gg R$ (see Fig.~\ref{fig:VortexDipoleEnergy}).
This interaction energy was already discussed in previous studies of Berezinskii-Kosterlitz-Thouless behavior for a thin cylindrical film \cite{Machta89}, and of vortex dipoles on capped cylinders \cite{Turn10}.
 
This analysis holds whenever a stream function describes the flow, even in the presence of boundaries when $\chi(\bm r,\bm r_j)$ involves two separate variables.  Here, it yields the general result 
 \begin{equation}\label{Egen}
E_{\rm tot} = -\frac{\pi n \hbar^2}{M}\sum_{j,k=1}^2 q_jq_k \chi(\bm r_j,\bm r_k),
\end{equation}
augmented by the  cutoff at $\xi$ when $j=k$.

 As seen below, this approach  also works for vortices on a finite cylinder, where the method of images gives the complex potential (see Sec.~\ref{sec:finiteCylinder}).   Finally, it describes the energy of point vortices in a planar annulus~\cite{Fett67} (see App.~\ref{sec:annulus}).
 
 \subsection{Vortex motion  as response to applied force}

 The modified interaction energy $V_{12}(\bm r_{12})$ in Eq.~(\ref{Vij}) allows us to rewrite 
the two vector dynamical equations near the start of Sec.~\ref{sec:induced_motion_of_two_vortices} as follows

\begin{equation}
q_1  \bm{\dot r}_1 = -\hat{\bm n} \times\bm\nabla_1 V_{12} \quad \hbox{and} \quad q_2  \bm{\dot r}_2 = -\hat{\bm n} \times\bm\nabla_2 V_{12}.  
\end{equation}
We can interpret the quantity $\bm F_1 =-\bm \nabla_1 V_{12}$ as the force that vortex 2 exerts on vortex 1, and similarly with $\bm F_2 = -\bm \nabla_2 V_{12} = -\bm F_1$, where the last relation follows because $V_{12}$ depends only on the difference of the coordinates.

In this way, the dynamical equations take the intuitive form (see Sec.~III of~\cite{Cald17})
\begin{equation}
q_1  \bm{\dot r}_1 =\hat{\bm n} \times\bm F_1 \quad \hbox{and} \quad q_2  \bm{\dot r}_2 = \hat{\bm n} \times\bm F_2 = - \hat{\bm n} \times\bm F_1.  
\end{equation}
Hence a vortex moves perpendicular to the applied force, which is often called  the Magnus effect.  Equivalently, we can introduce the ``Magnus force" $\bm F^M_j = q_j \hat{\bm n}\times \bm{\dot r}_j$, and the dynamical equation then  becomes $\bm F_j^M + \bm F_j = \bm 0$.  These equations concisely express two-dimensional vortex dynamics in a general form, applying not only to motion on a plane but also on a cylinder.

\subsection{Energy of multiple vortex dipoles}

As seen in Sec.~\ref{sec.iii},   the stream function for a set of $N$ vortices on an infinite cylinder is the sum of individual terms $\chi = \sum_{i=1}^N \chi_i$, where we assume $N$ is even.  The total kinetic energy of the vortices is proportional  to $\int d^2 r |\bm \nabla \chi|^2$ over the area of the cylinder.  This behavior is completely analogous to the electrostatic energy for two-dimensional point charges on a cylindrical surface, since the electrostatic energy is proportional to the integral $\int d^2 r|\bm E|^2$, and $\bm E $ is the (negative) gradient of the electrostatic potential $\cal P$.  Furthermore, the total electrostatic potential is a sum of contributions ${\cal P}_j$  from each charge, like the similar structure of the total $\chi$.  Finally, Eq.~(\ref{2d}) shows that the stream function $\chi$ obeys Poisson's equation with each vortex as a source, in complete analogy to the electrostatic potential $\cal P$ which also obeys Poisson's equation with the point charges as sources.

Thus, by analogy with two-dimensional electrostatics,  the energy of  multiple pairs of point vortices on the infinite cylinder follows immediately as the sum over all pairs plus the sum over all self energies
\begin{align}
\label{EtotVortexDipole}
E_{\rm tot} 
&= E_{\rm int} + E_{\rm self}= \sum_{i<j}^N E_{ij} + \sum_i^N E_i \nonumber\\
&=- \sum_{i<j}^N q_iq_j\,\frac{2\pi n\hbar^2}{M} \chi(\bm r_{ij}) + N \frac{\pi n \hbar^2}{M}\ln\left(\frac{2R}{\xi}\right).
\end{align}
If the system is overall neutral, then the total energy is finite; otherwise, there are divergent constant terms that do not affect the dynamics of individual vortices.  Similar divergences appear in two-dimensional electrostatics unless the total electric charge vanishes.

Equations (\ref{vFlow}), (\ref{vel}) and (\ref{EtotVortexDipole}) together give the general dynamical equations 
\beq
q_k\bm{\dot} x_k = \frac{\partial V_{\rm int}}{\partial y_k}\quad{\rm and}\quad q_k\bm{\dot} y_k =-\frac{\partial V_{\rm int}}{\partial x_k},
\eeq
where $V_{\rm int} = -\sum_{i<j}^N q_iq_j\,(\hbar/M) \chi(\bm r_{ij})$.  Thus $V_{\rm int} $ serves as a ``Hamiltonian''  with canonical variables $(x_k,y_k)$ that determines the motion of all the vortices.

  \section{Single vortex on a cylinder of finite length}\label{sec:finiteCylinder}

As seen in the previous sections, the complex potential generated by a single positive vortex located at the origin of a cylinder with radius $R$ and of infinite length, is
\begin{align}
F(z) = \ln \left[ \sin \left(\frac{z}{2R}\right)\right]. \label{InfiniteSolution}
\end{align}
The corresponding solution on a cylinder with finite length $L$ (with  $0\leq y \leq L$) follows with the method of images. 
Consider a physical vortex located at $z_0= (x_0+i y_0)$ with  $0 < y_0 < L$.  
Reflect the potential of the unbounded solution along the planes $y=i L$ and $y=0$ and reverse the charge of successive image vortices. 
This procedure creates an infinite set of positive vortices at positions $z_{(n,+)}=z_0+2inL$ and negative vortices at $z_{(n,-)} = z_0^*+2inL$.  We find
\begin{align} \label{SineProduct}
F_L(z) &= \sum_{n \in \mathbb{Z}}\left\{ \ln \left[ \sin \left(\frac{z-z_{(n,+)}}{2R}\right)\right]-\ln \left[ \sin \left(\frac{z-z_{(n,-)}}{2R}\right)\right]\right\} \nonumber \\[.2cm]
&= \ln \left[ \prod_{n \in \mathbb{Z}} \left( \frac{\sin(z_+/R - i\beta n )}{\sin(z_-/R - i\beta n )} \right) \right],
\end{align}
where $z_+ = (z-z_0)/2$, $z_-= (z-z_0^*)/2$, and $\beta = L/R$.

Examine the infinite product in Eq.\ \eqref{SineProduct} in detail:
\begin{widetext}
\begin{align}
\prod_{n \in \mathbb{Z}} \left( \frac{\sin(z_+/R - i\beta n)}{\sin(z_-/R - i\beta n )} \right) =& \frac{\sin(z_+/R)}{\sin(z_-/R)} \prod_{n =1}^{\infty} \left( \frac{\sin(z_+/R -i \beta n )\sin(z_+ /R+ i\beta n )}{\sin(z_- /R- i\beta n )\sin(z_-/R + i\beta n )} \right) \nonumber \\ 
=& \frac{\sin(z_+/R)}{\sin(z_-/R)} \prod_{n=1}^{\infty}\left( \frac{2\cos(2 z_+/R) - q^{2n}-q^{-2n}}{2\cos(2 z_-/R) - q^{2n}-q^{-2n}} \right) \nonumber \\
=& \frac{\vartheta_1(z_+/R , q)}{\vartheta_1(z_-/R, q)}, \label{SineProdAsTheta}
\end{align}
\end{widetext} 
where $q = e^{-\beta} = e^{-L/R}$.  Here, $\vartheta_1(z,q)$ denotes the first Jacobi $\vartheta$ function, defined by either its product form or its series form~\cite{Whit62}
\begin{align}
\vartheta_1(z , q) =& 2 q^{1/4} \sin(z) \prod_{n=1}^{\infty} (1-q^{2n})(1-2q^{2n}\cos(2z)+q^{4n}) \nonumber\\  
=&  2 \sum_{n=0}^{\infty} (-1)^n q^{(n+1/2)^2} \sin\left[(2n+1)z\right]. \label{ThetaSeries}
\end{align}
This function has simple zeros at the complex points $z = m\pi + n\pi\tau$, where $m,n \in \mathbb{Z}$ and $\tau$ is a complex number with positive imaginary part.  In addition, the parameter $q = e^{i\pi \tau}$ obeys the condition $|q|< 1$.  Here, $\tau = i\beta/\pi = iL/\pi R$ and hence $q=e^{-L/R}$, as noted above. 

\begin{figure*}[t]
\begin{center}
\includegraphics[width=\textwidth,angle=0,clip=]{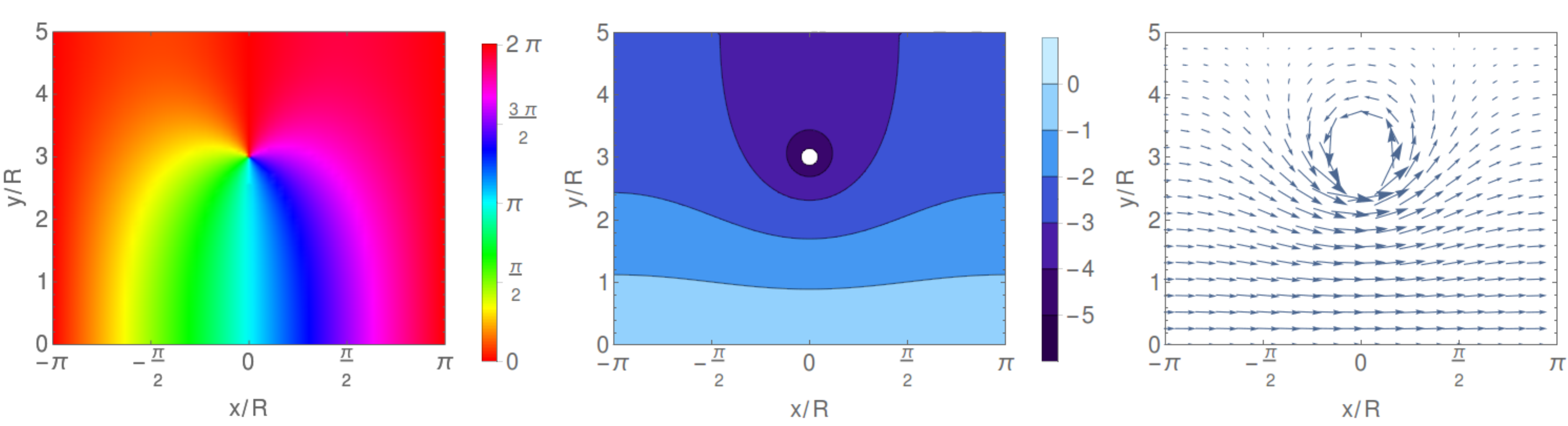}\end{center}
\caption{\label{fig:FlowCylFiniteLength}
Single vortex on a finite cylinder of length $L=5R$,  located at $y_0=3R$.
From left to right, plots show: the phase $\Phi$, the stream function $\chi$, and the velocity field $\bm {v}$.
} 
\end{figure*}

The final complex potential for a vortex located at $z_0$ on a cylinder of length $L$ and radius $R$ has the relatively simple analytic form 
\begin{align}
F_L(z) = \ln \left[\frac{\vartheta_1\left(\frac{z- z_0}{2R} , e^{-L/R}\right)}
{\vartheta_1\left(\frac{z-z_0^*}{2R} , e^{-L/R}\right)}\right].\label{PotentialFL}
\end{align}
Figure \ref{fig:FlowCylFiniteLength} shows the phase $\Phi(\bm r)= \Im F_L(z)$, the stream function $\chi(\bm r) = \Re F_L(z)$, and the vector velocity field $\bm v(\bm r)$ obtained from  $v_y + i v_x= (\hbar/M) F_L'(z)$.
These plots may be compared to the analogous ones for an infinite cylinder shown in Fig.~\ref{fig:singleVortexInfiniteCyl} (middle row).

The first Jacobi theta function $\vartheta_1(z,q)$ changes sign when $z \to z \pm \pi$, immediately proving that  the phase of the wave function changes by integer multiples of $2\pi$ when $x\to x\pm 2\pi R$. 
In particular, the integral $\int_{-\pi R}^{\pi R}{\rm d}x\;v_x$, computed at fixed $y$, equals 0 above the vortex $(y>y_0)$, and $2\pi \hbar/M$ below ($y<y_0)$, as is clear from Fig.~\ref{fig:FlowCylFiniteLength} (left).
Hence, the complex potential $F_L(z) $ always generates ``quantum-mechanically acceptable" solutions that  move uniformly around the cylinder.

By construction, the fluid is basically at rest above the vortex, and in motion below it.  This result may be understood by noting that the  original vortex and its image below the bottom end of the cylinder replicate a vortex dipole located at $(z_0,z_0^*)$.  In this basic ``building block", the fluid flow is largely confined to the region between the two vortices and vanishes at large distances from the line (or domain wall) joining the two vortices.

Note furthermore that $v_y$ is manifestly anti-symmetric around a vertical axis passing through the vortex core [namely, $v_y(x-x_0,y)=- v_y(x_0-x,y)$], so that its line integral along a circumference (at fixed $y$) vanishes. As a consequence, the angular momentum per particle on the cylinder is simply $\hbar(y_0/L)$. If angular momentum were to be ``pumped" at a constant (slow) rate into the system (namely, if the cylinder were to be spun with a linearly increasing rotation frequency, or if an increasing synthetic flux pierced the surface of the cylinder, as discussed in Ref.~\cite{Lacki2016}),  a vortex would enter the lower rim of the cylinder and progressively spiral up the cylinder. Once the vortex reaches the upper rim and leaves the cylinder, the angular momentum per particle would increase by exactly $\hbar$. This mechanism is  a direct hydrodynamic analog of the Laughlin pumping~\cite{Laug81}.

\subsection{Velocity of the vortex core}\label{ssv.1}
\begin{figure}[t]
\begin{center}
\includegraphics[width=.9\columnwidth,angle=0,clip=]{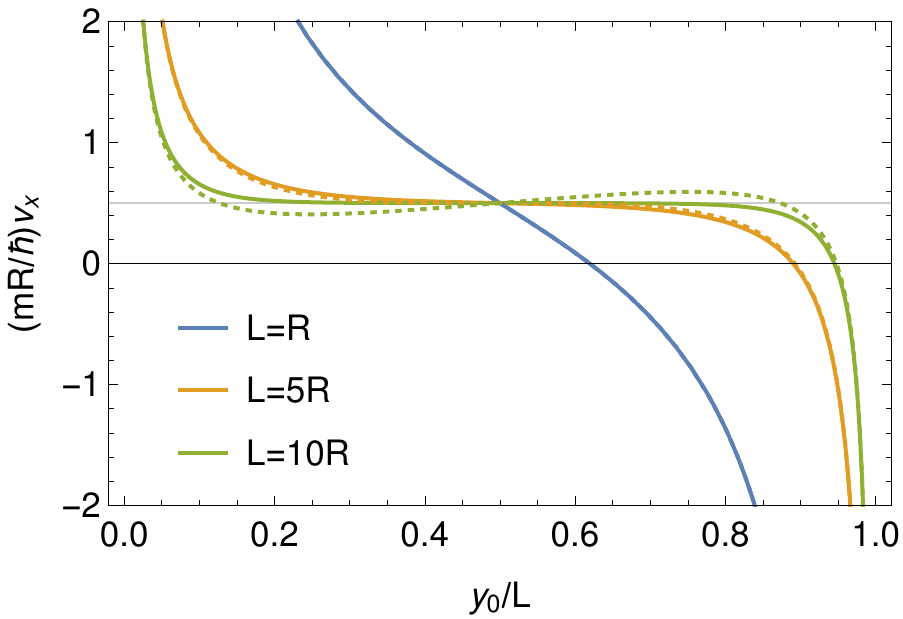}
\end{center}
\caption{\label{fig:vortexVelocity}
Velocity of the vortex core as a function of its vertical coordinate $y_0$, for various cylinder lengths $L$ (from top to bottom, on the left side, the lines represent $L=R,5R,10R$). The dotted lines are the approximate results, valid in the limit $L\lesssim \pi R$, obtained from Eq.\ \eqref{vxVortexApprox}. The blue dotted line is indistinguishable from the solid one. }
\end{figure}

The velocity of the vortex core follows from Eq.~(\ref{vortex_velocity}),
\beq\label{core_velocity}
\lim_{z\rightarrow z_0}\left(F'_L(z)-\frac{1}{z-z_0}\right)=-\frac{1}{2R}\frac{\vartheta'_1\left(i y_0/R , e^{-L/R}\right)}{\vartheta_1\left(i y_0/R , e^{-L/R}\right)},
\eeq
where $\vartheta'(z,q)$ indicates the derivative of $\vartheta$ with respect to its variable $z$.
This function is  purely imaginary,  indicating that the vortex moves solely along the $\hat{\bm x}$ direction, and its velocity diverges as it approaches either end  of the cylinder (namely, one of the image charges), as shown in Fig.\ \ref{fig:vortexVelocity}.  If the vortex is located at the middle of the cylinder at $z_0 = x_0 + i L/2$, one may use the property $\vartheta'_1(i u / 2 , e^{-u})=-i\vartheta_1(i u / 2 , e^{-u})$  (valid for generic real $u>0$) to show that it moves uniformly around the cylinder with speed 
\beq\label{speed_at_half_cylinder_length}
{\bm \dot x}_0 =\hbar/2MR\;\textrm{ when }\; y_0=L/2,
\eeq
 as seen in Fig.\ \ref{fig:vortexVelocity}.  Here the rightward motion arises because we chose to pair the vortex with its image in the lower boundary of the cylinder.  Had we instead used the image in the upper boundary at $z_0^* + 2iL$, the motion would  have been to the left with the same magnitude.  This broken symmetry is just that seen in Sec.\ \ref{secII} associated with the choice $C = \pm 1/2$.

\subsection{Analytical limits for long and short cylinders}
 
When $L\gg R$, the parameter $q$ is small, so that we may approximate 
$\vartheta_1(z,q) \approx  2 q^{1/4} \sin(z)$.  For a vortex at the complex position $z_0$, let $z\approx z_0 + z'$, where $z' = z-z_0$ is small.  Hence the previous $F_L(z)$ becomes
\begin{equation}
F_L(z) \approx \ln\left[\frac{\sin(z'/2R)}{\sin \left(2iy_0/2R  +z'/2R\right)}\right],
\end{equation}
where  we assume $z_0=iy_0$.  It is convenient to take $y_0 = L/2$, placing the vortex at the center of the cylinder.  

In this case, the denominator here becomes $\sin\left(iL/2R + z'/2R\right) \approx i\frac{1}{2} e^{L/2R}\left(1-iz'/2R\right)$.  As a result, we find the approximate expression
\begin{equation}
F_L(z) \approx \ln\sin\left(\frac{z'}{2R}\right) + \frac{iz'}{2R} - \frac{L}{2R} -\frac{i\pi}{2} + \ln 2.
\end{equation}
Apart from the additive constant terms, this result is precisely that found in Sec.~\ref{quantized} for an infinite cylinder with a vortex at the origin and zero velocity flow on its (distant)  upper rim.

When we have the opposite limit $L\ll R$, we may use the Jacobi imaginary transformation that relates a  theta function with parameter $\tau$ to one with parameter $\tau' = -1/\tau$~\cite{Whit62}. 
Here, $\tau' = i\pi R/L$ and $q'= e^{i\pi \tau'} = e^{-\pi^2R/L}$ is now small.
The relevant transformation formula becomes 
\beq\label{imag-trans}
\vartheta_1\left(\frac{z}{R} , e^{-L/R}\right) = i\sqrt{\frac{\pi R}{L}}\, e^{-z^2/RL}\,\vartheta_1\left(\frac{z\pi }{i L} , e^{-\pi^2 R/L}\right).
\eeq
In this way we find 
\begin{align}\label{FshortCyl}
F_{L\ll R}(z)&=\ln\left[
\frac{e^{-[\left(z-z_0\right)^2/4RL]}\,\vartheta_1\left(\frac{\pi (z-z_0)}{2iL}, e^{-\pi^2R/L}\right)}
{e^{-[\left(z-z_0^*\right)^2/4RL]}\,\vartheta_1\left(\frac{\pi(z-z_0^*)}{2iL}, e^{-\pi^2R/L}\right)}\right]\nonumber\\
&\approx
\ln\left[\frac{\sinh\left(\frac{\pi(z-z_0)}{2L}\right)} {\sinh\left(\frac{\pi(z-z_0^*)}{2L}\right)}\right]+i\frac{y_0}{L}\frac{z-x_0}{R}.
\end{align}
For a short cylinder, the result converges to the complex potential generated by a row of positive vortices located at positions $z_0+2imL$, together with a row of negative ones at positions $z_0^*+2imL$ (with $m\in \mathbb{Z}$). 

We use Eq.~\eqref{vortex_velocity} to find
the velocity of the vortex core on a short cylinder, 
\begin{align}\label{vxVortexApprox}
\frac{iM{\bm \dot x}_0}{\hbar} &=\lim_{z\rightarrow z_0}\left(F'_{L\ll R}(z)-\frac{1}{z-z_0}\right)\nonumber\\
&=-\frac{\pi}{2L}\coth\left(\frac{i \pi y_0}{L}\right)+i\frac{y_0}{LR}
=\frac{i \pi}{2L}\cot\left(\frac{\pi y_0}{L}\right)+i\frac{y_0}{LR}.
\end{align}
For a vortex at the center of the cylinder with $y_0 = L/2$, this equation gives the  familiar quantized  circulating motion ${\bm \dot x}_0 = \hbar/2MR$, in agreement with the result from Sec.~\ref{quantized} for $C= 1/2$.
In the limit $R\to \infty$, Eq.~(\ref{vxVortexApprox})  agrees with a result in Ref.\ \cite{Toikka2017}.

\subsection{Energy of a vortex dipole on a finite cylinder}

Although Eq.~(\ref{Egen}) gives the total energy of a vortex dipole on a finite cylinder, the following more physical approach clarifies the basic physics.  Consider a larger surface $-L\le y\le L$ which includes both the original vortex dipole and its vortex dipole image.  The superfluid flow is symmetric in $y$, so that the energy of this extended region is twice the original energy. The complex potential on a finite cylinder (and the corresponding stream function) may be decomposed in two contributions, one coming from the vortex itself, and one from the image vortex. The notation $\mathcal{L}(z)=\ln\left|\vartheta_1(z/2R,q)\right|$  denotes the part of the stream function due to the original vortex (and not to its image), and note that $\mathcal{L}(z)\approx\ln|\eta z/2R|$ for small $z$, with $\eta\equiv\vartheta_1'(0,q)$. To be very specific, the stream function obtained from $F_L(z)$ is $\chi(\bm r,\bm r_j) = {\cal L}(z-z_j) -{\cal L}(z-z_j^*)$.

In complete analogy with Eq.~\eqref{EtotVortexDipole}, the total energy on the extended cylinder due to the original vortex dipole and its image contains the core energies of the four vortices, given by the stream function $\mathcal{L}$ evaluated at the core radius, $E_{\rm core}\equiv-\frac{1}{2}[2\pi\mathcal{L}(\xi)]\approx\pi\ln(2R/\eta \xi)$, plus the stream function $\mathcal{L}$ evaluated for the relative separation of all six possible pairs of vortices. In particular, the energy on the extended cylinder due to a vortex dipole  with a positive vortex at $z_1=x_1+iy_1$ and a negative one at $z_2=x_2+ iy_2$ becomes
\begin{align}\label{Eext}
E_{\rm extended}=& 4E_{\rm core}-2\pi\sum_{i<j}q_iq_j\mathcal{L}(z_{ij})\\
=& 4\pi\ln\left(\frac{2R}{\eta \xi}\right)+2\pi\left[2\mathcal{L}(z_{12})+
\mathcal{L}(2 i y_{1})\right.\nonumber\\
&\left.+\mathcal{L}(2 i y_{2})-2\mathcal{L}\left( x_{12}+i y_1 +i y_2\right)\right].\nonumber
\end{align}
The energy of a vortex dipole on the original finite cylinder is $E_{\rm tot} = E_{\rm extended}/2$.

\section{Outlook and conclusions}

On a plane, a superfluid vortex represents a singularity.  The requirement that the condensate wave function $\Psi $ be single valued leads to the familiar  quantization of circulation around the vortex in units of $2\pi \hbar/M$.  A thin superfluid film on a cylindrical surface of radius $R$ allows for closed paths around the circumference of the cylinder as well as  those around a vortex.  As discussed in Sec.~\ref{quantized}, this condition requires that a single vortex on an infinite cylinder move in the azimuthal direction with uniform velocity $\pm \hbar/2MR$ as the simplest of many allowed quantized speeds.  Here, the choice of $\pm$ sign reflects a broken symmetry, corresponding to the two equivalent ``directions'' (up or down) along the axis of the cylinder.  Clearly, the topology of an infinite cylinder differs from that of a plane, here yielding these quantized translational  motions.

For a finite cylinder of length $L$, the boundary conditions require pairing with an image vortex in either end of the cylinder.  The resulting vortex dipole (the original  positive vortex and a negative image) automatically ensures that $\Psi$ is single valued and gives the appropriate translational velocity of the original vortex, where the choice of image (top or bottom) determines the broken symmetry and the sense of circumferential motion around the cylinder.

A superfluid film on the surface of a cylinder presents many experimental challenges.  Fortunately, this geometry is topologically equivalent to a superfluid on  a planar annulus, which was studied in detail in connection with three-dimensional superfluid $^4$He~\cite{Fett67,Bend62}.  There the interest was the sequence of equilibrium states as a function of the applied rotation.

More recently, the study of cold atoms has made dramatic progress in preparing superfluid annuli  with various dimensions~\cite{Ryu07, Moul12,Corm14,Eckel14,Vill17}, leading to the creation of very thin planar annuli with closely controlled radii~\cite{Aide17}.  To date, these recent measurements largely rely on interferometric techniques to study the superfluid velocity induced by various quenches.  

Earlier, rapid thermal quenches with three-dimensional condensates~\cite{Frei10}  created a single vortex line in  roughly 25\% of the events.  Furthermore  a clever technique allowed the study  of the vortex dynamics in real time, at intervals of $\sim 90$ ms.  
Alternatively, vortices may be created by merging multiple independent condensates \cite{Scherer07}.
We see no obvious reason why these methods cannot also apply to the study of a single vortex on a thin planar annulus. Appendix \ref{sec:annulus} examines the behavior of a single vortex on a planar annulus, using an inverse conformal transformation.

\section*{Acknowledgements} 
This work originated in a discussion with V.~S.~Bagnato, K.~Helmerson, W.~D.~Phillips, M.~Tsubota and A.L.F. at IIP in Natal, Brazil 
and continued during a visit by A.L.F. to ICFO in Barcelona.
The authors thank A.~Bachtold and M.~Lewenstein for many insightful discussions. 
We  also thank G.~Campbell and W.~D.~Phillips for extensive discussions about the NIST experiments on toroidal and annular condensates.
A.L.F. and P.M.  performed part of this work at the Aspen Center for Physics, which is supported by the National Science Foundation through Grant No.~PHY-1607611.
N.G. and P.M. acknowledge support from 
Spanish MINECO (Severo Ochoa SEV-2015-0522, 
and FisicaTeAMO FIS2016-79508-P), 
Generalitat de Catalunya (SGR 874, and CERCA), 
and the Fundaci\'o Privada Cellex.
N.G. is supported by a ``la Caixa-Severo Ochoa'' PhD fellowship.
P.M. further acknowledges funding from ``Ram\'on y Cajal" and ``Simons Foundation" fellowships.

\appendix
\section{Conformal mapping from the plane to the cylinder}\label{sec:conformal_transformation}

We review  briefly the elegant treatment of  Ho and Huang~\cite{Ho2015}, showing that their conformal transformation leads to results  that are compatible with ours.
Let  $Z=X+iY$ be the cartesian coordinates on the two-dimensional plane and  $z=x+iy$ the azimuthal and axial coordinates on the infinite cylinder (as in Sec.~\ref{secII}).
The phase pattern generated by a vortex located at position $Z_0$ in the plane is simply  $\Phi_{\rm plane}=\arg(Z-Z_0)$. The phase pattern on the cylinder is now simply $\Phi_{\rm cyl}=\arg(w-w_0)$, where $w$ is a conformal mapping from the plane to the cylinder.

Ho and Huang observe that there actually exist two (and only two) such mappings: $Z = w^{\pm} = e^{\pm i z}=e^{\pm i x}e^{\mp y}$. The map $w^+$ sends a tiny circle at the origin of the plane to the upper rim of the cylinder ($y\to \infty$), and a large circle on the plane to the lower rim of the cylinder ($y \to -\infty$), while $w^-$ does just the opposite.

As such, one vortex solution in the plane actually corresponds to two vortex solutions on the cylinder, with complex potentials 
\beq
F_{{\rm cyl},\pm}(z) = \ln\left( e^{\pm i z} - e^{\pm i z_0}\right)
\eeq
The corresponding phase patterns  are, respectively $\Phi_{\rm cyl,+}=\arg(e^{iz}-e^{iz_0})$, and $\Phi_{\rm cyl,-}=\arg(e^{-iz}-e^{-iz_0})$ (note that Ref.~\cite{Ho2015} uses $\Theta$ for the phase, instead of $\Phi$ as used here).
Setting $z_0=0$, we have
\beq
\Phi_{{\rm cyl},\pm}=\arg(e^{\pm iz}-1)={\rm Im}\{\ln[\sin(z/2)]\pm iz/2\}+\textrm{const.},
\eeq
which coincides with the result found in Eq.~(\ref{HH})
with a non-zero background flow set by  $C=\pm 1/2$, aside from an extra, irrelevant constant phase.
Similarly, $\ln|Z-Z_0|$ yields the stream function on the plane, so that $\ln|w-w_0|$ gives the stream function on the cylinder.

Equations~(9) and (10) in Ref.~\cite{Ho2015} give, respectively, the azimuthal and axial flow velocities $v_\phi$ and $v_z$ created by a vortex on the cylinder.
A straightforward calculation  shows that
\beq
\lim_{z\rightarrow z_0}\frac{\hbar}{M} \left[F'_{{\rm cyl,}\pm}(z)-\frac{1}{z-z_0}\right]=\pm i\frac{\hbar}{2MR},
\eeq
so  that the vortex core generated by the map $w^+$ ($w^-$) moves along the equator towards the right (left), with a velocity exactly equal to $\pm \hbar/(2MR)$, in  agreement with our conclusion at the end of Sec.~\ref{quantized}.

It is now simple to understand the ``need'' for (and amount of) quantization of the circulation around the top and bottom rims of the cylinder.
Focus, for example,  on the solution produced by $w^+$, and  consider a vortex in the plane, centered away from the origin ($Z_0\neq0$). A small circle around the origin of the plane encloses no vortices, and therefore has zero circulation, while a large circle always encircles the vortex, and therefore has circulation $2\pi$. As discussed above, the small and large circles are mapped by $w^+$ to the top and bottom parts of the cylinder, which indeed respectively show 0 and $2\pi$ circulations. (See the phase pattern in the central row of Fig.~\ref{fig:singleVortexInfiniteCyl}.)

\section{Single vortex on a planar annulus}\label{sec:annulus}

The surface of a cylinder of finite length is topologically equivalent to that of a planar annulus, and therefore we expect that the vortex dynamics will be very similar in the two cases. To derive the dynamics on the annulus, the simplest way to proceed is to apply the conformal mapping discussed above.

As seen in Sec.~\ref{sec:finiteCylinder}, the complex potential for a vortex located at $z_0$ on a cylinder of length $L$ and radius $R$ reads
\begin{align}
F_L(z) = 
\ln \left[\frac{\vartheta_1\left(\frac{z- z_0}{2R} , e^{-L/R}\right)}
{\vartheta_1\left(\frac{z-z_0^*}{2R} , e^{-L/R}\right)}\right]
+n_\uparrow\frac{i\,z}{R},\label{PotentialFL2}
\end{align}
where we have allowed for the possibility of having additional quantized flow on the upper rim of the cylinder, controlled by the integer number $n_\uparrow=C-1/2$. 

A convenient conformal mapping from the finite cylinder of radius $R$ to the annulus of radii $R_1=R_2\exp(-L/R)$ and $R_2$ is\beq
z=-i R\ln(Z/R_2),
\eeq
where $Z$ is the cartesian coordinate on the plane containing the annulus. This mapping sends the lower rim of the cylinder ($y=0$) to $R_2$, the upper rim ($y=L$) to $R_1$, and maintains the orientation, so that anti-clockwise rotation around the cylinder (increasing $x$) maps onto anti-clockwise rotation around the annulus (increasing polar angle).
Applying this mapping to Eq.~\eqref{PotentialFL2} immediately yields the potential for the annulus,
\begin{align}\label{F_annulus}
F_{\rm ann}(Z)&= F_{\rm circ}(Z)+F_{\rm images}(Z)\\\nonumber
&=
n_1\ln \left(\frac{Z}{R_2}\right) 
+\ln \left[\frac{\vartheta _1\left(-\frac{i}{2}\ln\left(\frac{Z}{Z_0}\right),\frac{R_1}{R_2}\right)}{\vartheta _1\left(-\frac{i}{2}\ln \left(\frac{Z Z_0^*}{R_2^2}\right),\frac{R_1}{R_2}\right)}\right],
\end{align}
where $n_1$ now determines the quantized flow circulation around the inner radius $R_1$ of the annulus \cite{footnote1967}. For the given mapping, we have simply $n_1=n_\uparrow$.

In the latter equation, the first term arises from the quantized flow around the inner boundary and the second arises from the images in both the inner and outer boundaries of the annulus, as can be shown by a direct calculation.
Indeed, if we include the original vortex and all the images, the positive vortices are at $ (R_1/R_2)^{2n}Z_0$ with $n\in \mathbb{Z}$. Correspondingly the negative vortices are at $(R_1/R_2)^{2n}Z_0^i$, where $Z_0^i$ denotes either  image $Z_0'= R_1^2/Z_0^*$ or $Z_0'' = R_2^2/Z_0^*$ and again  $n\in \mathbb{Z}$.  An analysis similar to that in Eq.~(\ref{SineProduct}) reproduces the previous expression $F_{\rm images}(Z)$, but now with either image, which makes clear the symmetry between the inner radius $R_1$ and the outer radius $R_2$.  For the image in the inner (outer) radius, the vortex precesses in the negative (positive) direction.

\begin{figure*}[t]
	\begin{center}
		\includegraphics[width=\textwidth,angle=0,clip=]{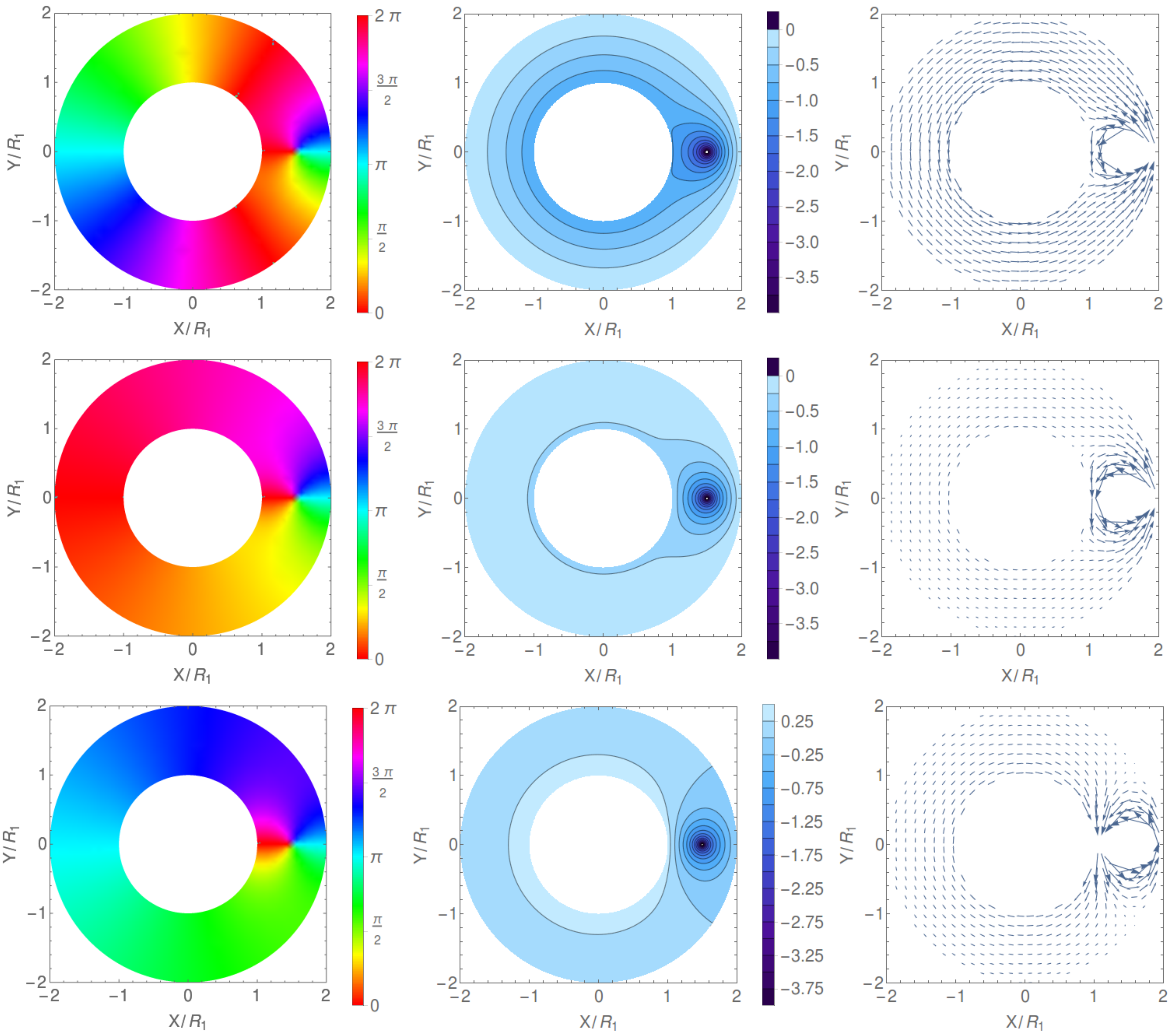}
	\end{center}
	\caption{\label{fig:FlowAnnulus}
		Single vortex on an annulus: (top) $n_1=1$, (center) $n_1=0$, and (bottom) $n_1=-1$. In all cases, the vortex is located at $z_0=1.5R_1$, and the outer radius is $R_2=2R_1$.
		From left to right, plots show: the phase $\Phi$, the stream function $\chi$, and the velocity flow.}
\end{figure*}
\begin{figure*}[t]
	\begin{center}
		\includegraphics[width=.45\textwidth,angle=0,clip=]{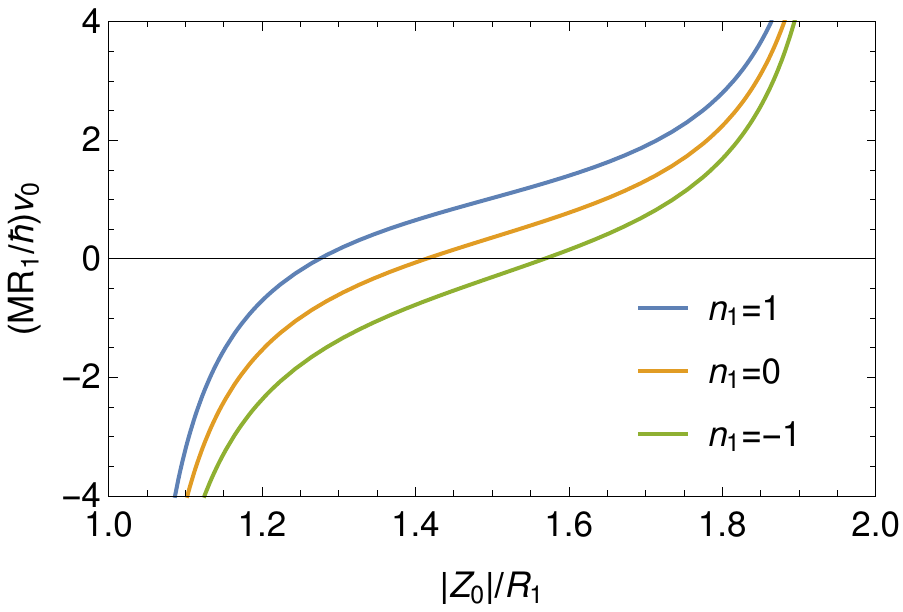}
		\hspace{1cm}
		\includegraphics[width=.45\textwidth,angle=0,clip=]{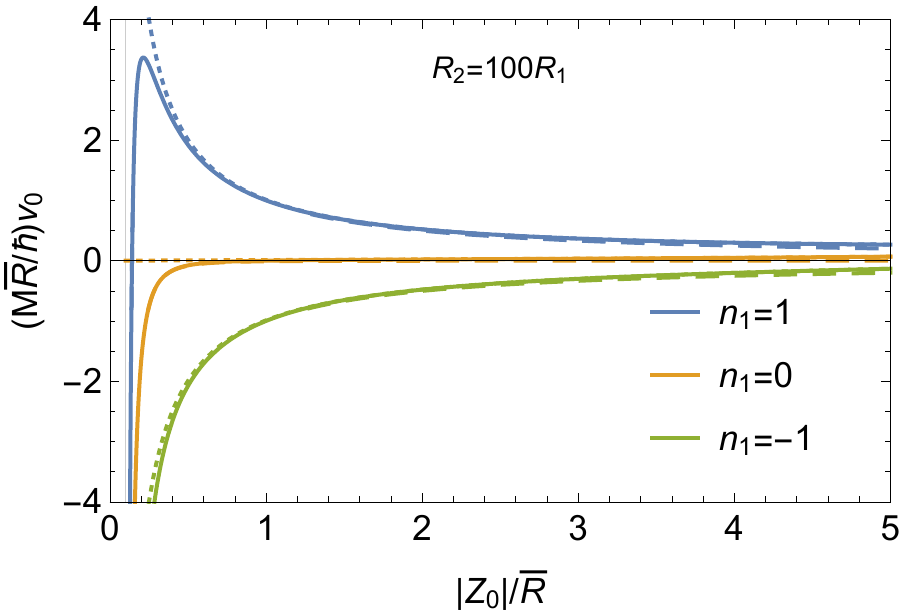}
	\end{center}
	\caption{\label{fig:vortexVelocityAnnulus}
		{ Velocity of the vortex core on an annulus.} Left: results for an outer radius $R_2=2R_1$, and for various values of the flow circulation around the inner radius (from top to bottom, $n_1=1,0,-1$). 
		Right: wide annulus limit ($R_2\gg R_1$); $\bar{R}=\sqrt{R_1 R_2}$ is the geometric mean of the radii, and the dotted and dashed lines show the limiting cases $R_1\rightarrow 0$ and $R_2\rightarrow\infty$, respectively given by Eq.~\eqref{tangVelocityPuncturedCircle} and \eqref{tangVelocityPlaneWithAHole}.}
\end{figure*}

The velocity potential $\Phi$, the stream function $\chi$ and the velocity flow are shown in Fig.~\ref{fig:FlowAnnulus} for $n_1=0$ and  $n_1=\pm1$.
The plots for non-zero circulation $n_1=\pm1$ clearly indicate that the multiply connected geometry of the planar annulus (and of the cylinder) allows for the presence of two distinguishable and independent phase windings: the one around the vortex, and the one around the inner boundary.

A vortex on a planar  annulus precesses around the center of the system, similar to the case of the cylindrical surface.
A detailed calculation gives the tangential (precessional) velocity of the vortex core
\beq\label{tangVelocity}
v_0
=\frac{\hbar}{M |Z_0|}\left[n_1 -\frac{1}{2} +
\frac{i}{2}\frac
{\vartheta_1'\left(-i\ln\left(\frac{|Z_0|}{R_2}\right),\frac{R_1}{R_2}\right)}
{\vartheta_1\left(-i\ln\left(\frac{|Z_0|}{R_2}\right),\frac{R_1}{R_2}\right)}
\right].
\eeq
To derive this result one needs to note that, in the vicinity of the vortex core $Z_0$,
\begin{multline}
Q(Z)\equiv-\frac{i}{2Z}\frac{\vartheta_1'\left(-\frac{i}{2}\ln\left(\frac{Z}{Z_0}\right),\frac{R_1}{R_2}\right)}
{\vartheta_1\left(-\frac{i}{2}\ln\left(\frac{Z}{Z_0}\right),\frac{R_1}{R_2}\right)}
\approx
\frac{1}{Z-Z_0}-\frac{1}{2Z_0},
\end{multline}
as may be seen expanding the logarithms inside the $\vartheta$ functions to {\it second order} in $Z-Z_0$, so that $\lim_{Z\rightarrow Z_0} \left(Q(Z)-\frac{1}{Z-Z_0}\right)=-\frac{1}{2Z_0}$.

When $|Z_0|=\bar{R}\equiv\sqrt{R_1 R_2}$ (the geometric mean of the inner and outer radii), the identity $\vartheta_1'\left(-i\ln\sqrt q,q\right)=-i \vartheta_1\left(-i\ln\sqrt q,q\right)$ valid for any real $q$ ($0<q<1$) immediately yields the simple result for the precessional velocity  
\beq
v_0 =n_1(\hbar/M \sqrt{R_1R_2}).
\eeq
It is intriguing to note that the mapping transforms the circle of radius $\bar{R}$ onto the circle $y=L/2$ going round a finite cylinder at half its length. As such, the latter result is the direct analog of Eq.~\eqref{speed_at_half_cylinder_length}.  Note, however, there is an important difference: when $n_1=n_\uparrow=0$ a vortex along this line is stationary on the annulus, but it moves on the cylinder.

The series expansion of $\vartheta_1$ for small $q$ may now be used: $\vartheta_1(\alpha,q)=2q^{1/4}(\sin\alpha-q^2\sin3\alpha)+{\cal O}(q^{25/4})$.
Retaining only the lowest order, we find the limiting behavior for $R_1\rightarrow 0$,
\beq\label{tangVelocityPuncturedCircle}
v_0
=\frac{\hbar}{M |Z_0|}\left(n_1 + \frac{|Z_0|^2}{R_2^2-|Z_0|^2}\right).
\eeq
When $n_1=0$, one recovers the well-known result for a vortex in a trapped disk-shaped BEC. 

To take the limit $R_2\rightarrow\infty$ while retaining a dependence on $|Z_0|$, one needs instead to expand both $\vartheta_1$ functions to  order $q^{9/4}$. Doing so gives a rather complicated expression.   In the limit $R_2\rightarrow \infty$  we find
\beq\label{tangVelocityPlaneWithAHole}
v_0
=\frac{\hbar}{M |Z_0|}\left(n_1 - \frac{R_1^2}{|Z_0|^2- R_1^2}\right),
\eeq
which is the usual result for a vortex outside a cylinder of radius $R_1$.  Finally in the limit $R_1 \ll |Z_0| \ll R_2$, one finds $v_0 = n_1 \hbar/M|Z_0|$.

Figure \ref{fig:vortexVelocityAnnulus} shows the various results found above for the precession velocity of a vortex  on an annulus. 
A confirmation of the validity of the above findings may be obtained by showing that the complex potential for the annulus with $n_1=0$ reduces to the one for the disk when $R_1\rightarrow 0$. Indeed, using the lowest order of the series expansion of the theta function, simple algebra shows that
\begin{eqnarray}
F_{\rm ann}(Z)&\approx
\ln \left[\frac{\sin\left[-\frac{i}{2}\ln\left(\frac{Z}{Z_0}\right)\right]}{\sin\left[-\frac{i}{2}\ln \left(\frac{Z Z_0^*}{R_2^2}\right)\right]}\right]\nonumber\\
&=\ln\left[\frac{Z-Z_0}{Z-\frac{R_2^2}{Z_0^*}}\right]+
\textrm{ a const.}
\end{eqnarray}
Since the complex potential for the annulus reduces to the one of the disk (apart from an irrelevant constant), the velocity must do the same.


\begin{thebibliography}{99}
\bibitem{footnote1967}
Using the Jacobi imaginary transformation, the complex potential for the annulus may be written as 
\begin{multline}
F_{\rm ann}(Z)=n_1\ln \left(\frac{Z}{R_2}\right)+ 
\frac{\ln \left(\sqrt{\frac{Z_0^*}{Z_0}}\frac{Z}{R_2}\right)\ln \left(\frac{R_2}{\left| Z_0\right| }\right)}{\ln\left(\frac{R_2}{R_1}\right)}\\
+\ln \left[\frac{\vartheta _1\left(\gamma\ln\left(\frac{Z}{Z_0}\right),q\right)}{\vartheta _1\left(\gamma  \ln \left(\frac{Z Z_0^*}{R_2^2}\right),q\right)}\right],\nonumber
\end{multline}
where $\gamma=\pi/[2\ln(R_2/R_1)]$ and $q=e^{-2\pi\gamma}$. This formula differs from the one provided in Ref.~\cite{Fett67} by the phase factor $\sqrt{Z_0^*/Z_0}$, which represents a rotation of the branch cut of the logarithm. Whenever $Z_0$ lies on the real positive axis, however, the phase factor is unity and the two formulas coincide. Note that the roles of $\chi$ and $\Phi$ are interchanged with respect to Ref.~\cite{Fett67}, so that we have removed an $i$ from Eq.~(10) of that work.
\end{thebibliography}
\end{document}